\documentclass[conference,10pt,svgnames]{IEEEtran}

\usepackage[switch]{lineno}
\usepackage{cite}
\usepackage{subfig}
\usepackage{url}
\usepackage[hidelinks]{hyperref}
\usepackage{enumitem}

\makeatletter
\begingroup\endlinechar=-1\relax
       \everyeof{\noexpand}%
       \edef\x{\endgroup\def\noexpand\homepath{%
         \@@input|"kpsewhich --var-value=HOME" }}\x
\makeatother

\usepackage{import}
\usepackage{etoolbox}
\newrobustcmd{\toolspath}[0]{latex-tools-overleaf/}

\usepackage{etoolbox}

\newtoggle{AlgsNoEnd}

\providebool{tr}
\providebool{conf}

\usepackage[utf8]{inputenc}
\newcommand{\code}[1]{\texttt{#1}}
\usepackage[binary-units=true]{siunitx}
\usepackage{xspace}
\usepackage[svgnames]{xcolor}

\usepackage{collab}
\usepackage[numbers,sort&compress]{natbib}

\usepackage{soul}
\soulregister\cite7
\soulregister\autoref7
\soulregister\code7
\soulregister\toolname7
\soulregister\ref7
\soulregister\emph7
\soulregister\pageref7
\usepackage{hhline}
\definecolor{lightyellow}{RGB}{250, 250, 180}
\definecolor{lightblue}{RGB}{145, 183, 242}
\definecolor{HLYELLOW}{RGB}{240, 127, 0}
\definecolor{hlyellow}{RGB}{240, 127, 0}
\sethlcolor{lightyellow}

\usepackage{algorithm}
\usepackage[noend]{algpseudocode}
\usepackage{pifont}
\usepackage{amsmath}
\algdef{SE}[FOREACHIN]{ForEachIn}{EndForEachIn}[2]{\algorithmicfor\ \textbf{each}\ #1\ \textbf{in}\ #2\ \textbf{do}}{\algorithmicend\ \algorithmicfor}%
\algdef{SE}[FOREACHP]{ForEachP}{EndForEachP}[1]{\algorithmicfor\ \textbf{each}\ #1\ \textbf{in parallel do}}{\algorithmicend\ \algorithmicfor}%
\algdef{SE}[FOREACHINP]{ForEachInP}{EndForEachInP}[2]{\algorithmicfor\ \textbf{each}\ #1\ \textbf{in}\ #2\ \textbf{in parallel do}}{\algorithmicend\ \algorithmicfor}%
\algdef{SE}[FORMOD]{ForMod}{EndForMod}[3]{\algorithmicfor\ #1\ \textbf{from}\ #2\ \textbf{to}\ #3\ \textbf{do}}{\algorithmicend\ \algorithmicfor}%
\algdef{SE}[FORMOD]{ForModS}{EndForModS}[4]{\algorithmicfor\ #1\ \textbf{from}\ #2\ \textbf{to}\ #3\ \textbf{step}\ #4\ \textbf{do}}{\algorithmicend\ \algorithmicfor}%
\algdef{SE}[FORMODP]{ForModPS}{EndForModPS}[3]{\algorithmicfor\ #1\ \textbf{from}\ #2\ \textbf{to}\ #3\ \textbf{in parallel do}}{\algorithmicend\ \algorithmicfor}%
\algdef{SE}[FORMODP]{ForModP}{EndForModP}[4]{\algorithmicfor\ #1\ \textbf{from}\ #2\ \textbf{to}\ #3\ \textbf{step} #4\ \textbf{in parallel do}}{\algorithmicend\ \algorithmicfor}%

\iftoggle{AlgsNoEnd} {
  \algtext*{EndFunction}% Remove "end while" text
  \algtext*{EndIf}% Remove "end if" text
  \algtext*{EndFor}% Remove "end if" text
  \algtext*{EndForEachIn}% Remove "end if" text
  \algtext*{EndForEachP}% Remove "end if" text
  \algtext*{EndForEachInP}% Remove "end if" text
  \algtext*{EndForModP}% Remove "end if" text
  \algtext*{EndForModPS}% Remove "end if" text
  \algtext*{EndForModS}% Remove "end if" text
}{}

\algdef{SE}[VARIABLES]{Variables}{EndVariables}{\algorithmicvariables}
  {\algorithmicend\ \algorithmicvariables}
\algnewcommand{\algorithmicvariables}{\textbf{global}}
\algnewcommand{\LineComment}[1]{\State \(\triangleright\) #1}
\algnewcommand{\And}{\textbf{and}\xspace}

\usepackage{tikz}
\usepackage{pifont}
\usetikzlibrary{shapes}

\DeclareRobustCommand*\circledColorSmall[2]{\tikz[baseline=(char.base)]{
    \node[shape=circle,fill=#2,draw=#2,inner sep=0pt] (char) {\textcolor{white}{\footnotesize\textbf{#1}}};}}

\usepackage{float}
\usepackage{listings}
\newfloat{codeblock}{H}{myc}
\definecolor{darkblue}{rgb}{0,0,.6}
\definecolor{darkred}{rgb}{.6,0,0}
\definecolor{darkgreen}{rgb}{0,.5,0}
\definecolor{red}{rgb}{.98,0,0}
\definecolor{gray}{rgb}{.6,.6,.6}
\definecolor{newgreen}{RGB}{169,209,142}
\definecolor{newpurple}{RGB}{237,134,254}
\definecolor{neworange}{RGB}{244,177,131}
\definecolor{newyellow}{RGB}{255,217,102}
\collabAuthor{copik}{blue}{Marcin}
\collabAuthor{lex}{violet}{lex}
\collabAuthor{htor}{magenta}{Torsten}

\usepackage{booktabs}
\usepackage{multirow}
\usepackage{adjustbox}
\usepackage{moresize}

\newcommand{\toolname}{\emph{rFaaS}\xspace}

\definecolor{newBlue}{RGB}{162, 210, 255}

\definecolor{newRed}{RGB}{199, 58, 48}

\usepackage{adjustbox}
\newenvironment{bluebox}{%
\noindent
\adjustbox{innerenv={varwidth}[c]{0.9\linewidth},margin=\fboxsep+.25cm \fboxsep,bgcolor=newBlue!60,center}\bgroup
}{%
\egroup
}

\usepackage{tcolorbox}

\usepackage{xcolor}
\usepackage{soul}
\usepackage{marginnote}

\author{
    \IEEEauthorblockN{
        Marcin Copik\IEEEauthorrefmark{1},
        Marcin Chrapek\IEEEauthorrefmark{1},
        Larissa Schmid\IEEEauthorrefmark{2},
        Alexandru Calotoiu\IEEEauthorrefmark{1},
        Torsten Hoefler\IEEEauthorrefmark{1}
    }
    \IEEEauthorblockA{\IEEEauthorrefmark{1}Department of Computer Science,
      ETH Z{\"u}rich, Z{\"u}rich, Switzerland}
    \IEEEauthorblockA{\IEEEauthorrefmark{2}Karlsruhe Institute of Technology, Germany}
    \IEEEauthorrefmark{1}firstname.lastname@inf.ethz.ch
    \IEEEauthorrefmark{2}larissa.schmid@kit.edu
}

\begin{document}

%\linenumbers

\title{Software Resource Disaggregation for HPC with Serverless Computing}

\maketitle

\begin{abstract}
Aggregated HPC resources have rigid allocation systems and programming models
which struggle to adapt to diverse and changing workloads.
Consequently, HPC systems fail to efficiently use the large pools of unused memory and increase the utilization
of idle computing resources.
%the large pools of idle memory.
%
%
Prior work attempted to increase the throughput and efficiency of supercomputing systems through workload co-location and resource disaggregation.
However, these methods fall short of providing a solution that can be applied to existing
systems without major hardware modifications and performance losses.
In this paper, we improve the utilization of supercomputers by employing the new cloud paradigm of serverless computing.
We show how serverless functions provide fine-grained access to the resources of batch-managed cluster nodes.
We present an HPC-oriented Function-as-a-Service (FaaS) that satisfies the requirements of high-performance applications.
%
%We show that the Function-as-a-Service (FaaS) programming model satisfies the requirements of high-performance applications
%and how idle memory helps resolve cold startup issues.
%
We demonstrate a \emph{software resource disaggregation} approach where placing functions on unallocated and underutilized nodes allows idle cores and accelerators to be utilized while retaining near-native performance.

\end{abstract}

\noindent \textbf{HPC FaaS Implementation}: \url{https://github.com/spcl/rFaaS}

\section{Introduction}

% idle nodes
% memory service
% focus on rfaaS

\begin{figure}[t!]
	\centering
    %\resizebox*{\width}{0.5\totalheight}{
    \subfloat[Idle CPU cores rate (\%).]{%	
      \includegraphics[width=\linewidth]{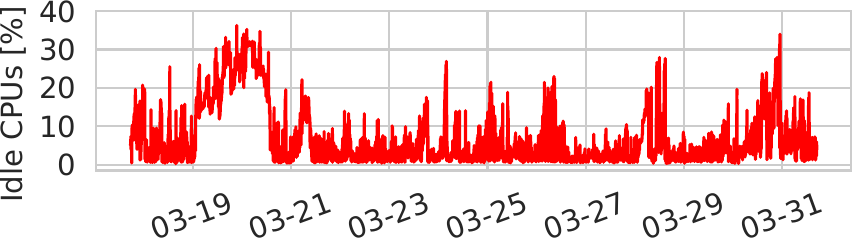}
      \label{fig:daint_cpu}
    }
	%}
	%\hfill
	
 %\resizebox*{0.45\width}{0.5\totalheight}{
		\subfloat[Free memory rate (\%).]{%	
      \includegraphics[width=\linewidth]{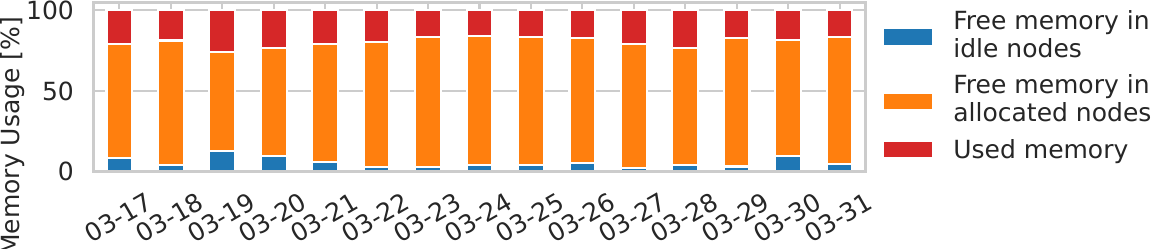}
      \label{fig:daint_memory}
		}

		\subfloat[Duration of idle periods for nodes. Estimation from discrete sampling.]{%	
      \includegraphics[width=\linewidth]{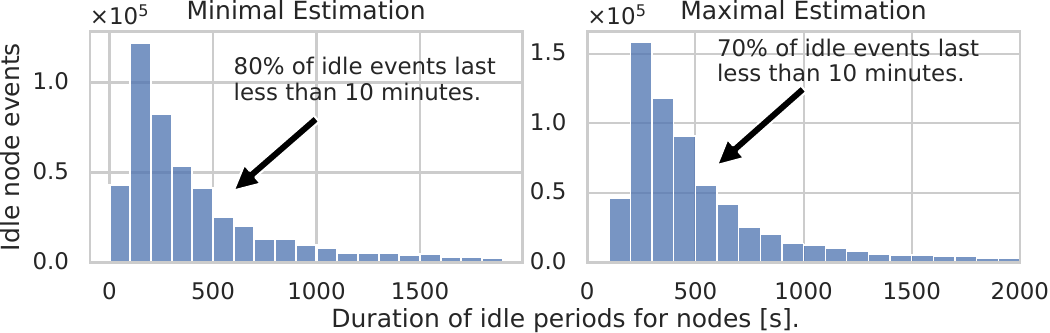}
      \label{fig:daint_idle}
		}
  
		%\subfloat[Duration of idle periods for nodes.]{%	
      %\includegraphics[width=\linewidth]{figures/nodes_idle_statistics}
      %\label{fig:daint_idle}
		%}
	%\hfill
 
	%\subfloat[Duration of idle periods for nodes.]{%	
    %  \includegraphics[width=\linewidth]{figures/nodes_idle_statistics}
    %  \label{fig:daint_idle}
	%	}
  \caption{Piz Daint utilization in March 2022: querying SLURM with a two-minute interval. See Sec.~\ref{sec:background_utilization_hpc} for details.}
  %querying SLURM with a two minute interval.}
  \label{fig:daint_stats}
\end{figure}

% how does the landscape look like?
% need to disaggregate
% diverse resources
% challenges - programming model and co-location
%
Modern HPC systems come in all shapes and sizes, with varying computing power, accelerators,
memory size, and bandwidth~\cite{10.1145/3432261.3432263}.
Yet, most of them share one common characteristic: resource underutilization.
Past predictions showed a pessimistic research outlook:
\emph{"the goal of achieving near 100\% utilization while supporting a real parallel supercomputing workload is unrealistic"}~\cite{10.1007/3-540-47954-6_1}.
Node utilization of supercomputer capacity varies between 80\% and 94\% on different systems~\cite{10.1007/978-3-642-35867-8_14,10.5555/3433701.3433812,DBLP:journals/corr/JonesWIDSPGFSBK17}, with
up to 75\% of memory is underutilized as these resources are overprovisioned for
workloads with the greatest demands (Fig.~\ref{fig:daint_stats}).
A 10\% decrease in monthly utilization can lead to hundreds of thousands of dollars
of investment in unused hardware.
This gap cannot be addressed with persistent and long-running allocations.
HPC operators should incentivize users to use spare CPU cores or idle GPUs to accelerate their applications,
improving the cost and energy efficiency of the system.
To that end, users need \emph{fine-grained resource allocations}
and \emph{elastic programming models}.

In an HPC system, wasted resources are found in idle and allocated nodes.
Most idle nodes are inactive only for several minutes (Fig.~\ref{fig:daint_idle}) and cannot be integrated into long-running and static batch allocations.
On the other hand, hardware can remain idle on an allocated node due to overprovisioning and a mismatch between available resources and job demands; even the optimal number of threads is application-specific and
\emph{”rarely equal to the number of cores on the processor”}~\cite{https://doi.org/10.1002/cpe.3187}.
\emph{Evolving} and \emph{malleable} applications cannot adjust resource allocation in rigid HPC systems~\cite{10.1007/BFb0022284},
leading to severe underutilization, as runtime adaptivity could reduce core-hour consumption by up to 44\% in malleable applications~\cite{10.1145/3075564.3075585,8026084}.
The wasted on-node resources cannot be employed by another application due to the coarse granularity of batch allocations in HPC.
In data centers, the problems of underutilization and coarse-grained allocations are resolved with techniques such as \emph{resource disaggregation} and \emph{job co-location}.
However, the tight coupling of resources and performance constraints make their direct application to HPC systems difficult.

%The former achieves that by consolidating resources allocated later in a fine-grained
%manner in the exact amount needed by the application.
%
%At the same time, the latter improves utilization by mixing workloads with compatible resource consumption patterns.
%
%Still, neither of them has seen a major adoption in HPC yet.

% Resource disaggregation refers to the ability of a system to pool and compose resources in a ine-grain manner
% and thus be capable of allocating exactly the resources an application requests. This is in contrast to many systems
% today where nodes are allocated to applications as a unit with identical ixed-sized resources; any resources
% inside nodes that the application does not use have no choice but to idle. Following the trend for hardware
% specialization and the desire to better utilize resources as systems scale up, resource disaggregation across the
% system or a group of racks has been actively researched and deployed in commercial hyperscale datacenters in
% Google, Facebook, and others [17, 55, 64]. In addition, many studies focus on disaggregation of GPUs [35] and
% memory capacity [33, 67].
% ~\cite{10.1145/3514245}.
%
% This paper is a good overview: https://dl.acm.org/doi/abs/10.1145/3127479.3131612

%The former achieves that by consolidating resources allocated later in a fine-grained
%manner in the exact amount needed by the application.
Disaggregation improves utilization by consolidating resources and allocating later in the amount needed by the application (Sec.~\ref{sec:background_resource_disaggregation}).
Disaggregation targets specialized hardware~\cite{10.1145/3514245}
and improves memory's performance--per--dollar by up to 87\%~\cite{10.1145/1555754.1555789}.
Memory disaggregation can be supported in hardware~\cite{9252003,10.1145/3127479.3131612,10.1145/1555754.1555789},
but these solutions require dedicated extensions and have high costs~\cite{10.1145/3127479.3131612}.
%
%However, remote memory comes with scalability and fault tolerance challenges~\cite{10.1145/3127479.3131612}.
%communication needed to access remote resources impacts bandwidth
%However, the communication needed to access remote resources impacts bandwidth
%and latency~\cite{10.1145/3514245,9235066} and remote memory comes with
%scalability and fault tolerance challenges~\cite{10.1145/3127479.3131612}.
%
Instead, we propose a software system that capitalizes on available high-performance interconnects and \textbf{runs on the HPC systems existing today}.

%"Resource utilization and through-
%put suffers when small jobs are forced to occupy an entire node while making use of
%only a few processors. The policy also encourages users to squeeze large parallel jobs
%onto the fewest number of nodes possible since doing otherwise is both costly and detri-
%mental to system utilization. Such configurations are not always optimal;the processes of parallel jobs often perform similar computations, consequently stressing the same
%shared resources and exacerbating the slowdown due to resource contention."
%~\cite{10.1007/978-3-540-71035-6_10}
% "Enabling Fair Pricing on HPC Systems with Node Sharing”
% "Reducing network contention with mixed workloads on modernmulticore, clusters."
% "The case for colocation of hpc workloads."
% Oversubscription on multicore processors
While sharing HPC nodes by co-located jobs can improve performance and efficiency~\cite{5470434,5289162},
%
%However, assigning CPU cores to each job is insufficient since applications sharing other resources on a node
%can cause interference.
%
% Applicatons on different socket can suffer from interference via the underlying processor interconnect~\cite{6468475}.
%
space-sharing by applications that simultaneously stress the same resources will lead to contention~\cite{10.5555/2388996.2389109,6468475}.
%
%Co-located applications - memory contention between 10-60\% slowdown and
%"up to several orders of magnitude for I/O~\cite{10.1007/978-3-540-71035-6_10}
%Memory bus saturation causes up to 3-fold decrease in performance of bandwidth consuming applications~\cite{1240622}.
%Bandwidth is a precious resource in supercomputing systems - the bandwidth-per-flop ratio has been
%decreasing for the last decade~\cite{10.1145/3432261.3432263}.
%
Memory and I/O contention cause a slowdown of up to three times and several orders of magnitude, respectively~\cite{10.1007/978-3-540-71035-6_10,1240622,10.1145/3432261.3432263},
and many systems disable node sharing for that reason.
%
%While the disabling of node sharing might seem to promise performance stability and predictability and exclusive access to hardware,
%the HPC applications still suffer from hardware contention and interference
%in the shared high-speed interconnect~\cite{7161533,6877474} and parallel filesystem~\cite{7516071}.
%This is important - up to 35\% of application time can be sped on message-passing communication~\cite{4548205}.
%
%Thus, many supercomputing systems disable node sharing for that reason, even though it does not provide
%performance stability and predictability - applications still suffer from contention in the
%high-speed interconnect~\cite{7161533,6877474} and parallel filesystem~\cite{7516071}.
%
%While \emph{job striping}~\cite{https://doi.org/10.1002/cpe.3187,5289162} increases performance
%by spreading application processes and co-locating them with other workloads,
%it requires understanding the \emph{symbiosis} of co-located applications
%or \emph{partitioning} shared sources (Sec.~\ref{sec:background_colocation}).
To reduce interference, users and system operators have to understand the \emph{symbiosis} of co-located workloads (Sec.~\ref{sec:background_colocation}).
Furthermore, new approaches to security are needed when sharing bare-metal HPC nodes.

We target unused resources on idle and allocated nodes by bringing the flexibility and isolation of cloud abstraction models to HPC.
We propose to use the \emph{Function--as--a--Service (FaaS)} programming model, where users invoke fine-grained and short-running functions.
Invocations are executed by the system operator on dynamically provisioned resources in a \emph{serverless} fashion.
%
%The cloud provider handles function invocations on dynamically
%allocated resources in a \emph{serverless} fashion. 
%Sec.~\ref{sec:background_faas}).
FaaS introduces three major improvements that make it suitable for HPC resource management:
%
%(1) Since functions are time-limited, a temporarily available node can handle invocations and still be quickly drained for batch jobs;
%
%(2) Functions can be co-located on the same node, where each one is using a different set of resources and is fully isolated from others;
%
%(3) In FaaS, the system operator has full control over function placement, and can thus opportunistically allocate invocations to fill utilization gaps left by a batch job executing on overprovisioned hardware.
\setlength{\parskip}{0pt}
\setlength{\parsep}{0pt}
\begin{itemize}[noitemsep,topsep=0pt]
\item A temporarily available node can handle time-limited functions and still be quickly drained for batch jobs;
\item Functions can be co-located on the same node, where each one is using a different set of resources and is fully isolated from others;
\item In FaaS, the system operator has full control over function placement, and can thus opportunistically allocate invocations to fill utilization gaps left by a batch job executing on overprovisioned hardware.
\end{itemize}
%
%The fine-grained allocations with \emph{pay--as--you--go} billing could resolve the problem of runtime adaptivity in HPC.
%
%Yet, no work has fully embraced this cloud revolution to improve the efficiency
%of existing supercomputers.
%
%The question of incorporating elastic cloud resources into MPI applications
%remains open due to a lack of a cloud-native programming model.

\begin{figure}[!t]
	\centering
  \includegraphics[width=1.0\linewidth]{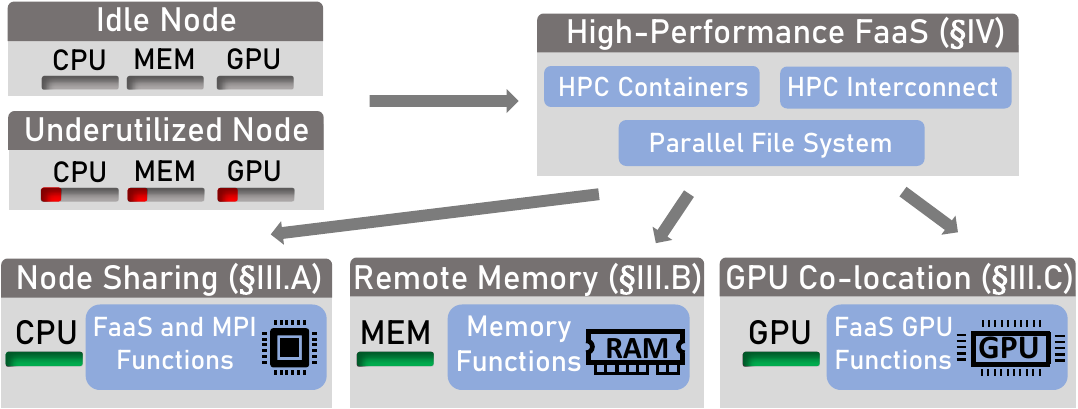}
  \caption{\textbf{Software disaggregation with FaaS}: increasing resource utilization
  without modifications to HPC hardware.}
  \label{fig:intro_overview}
\end{figure}

In this paper, we present the first FaaS system
that implements \textbf{software disaggregation} of resources in a supercomputing system  (Fig.~\ref{fig:intro_overview}).
We show that dynamic function placement provides a functionally equivalent solution to
disaggregated computing on homogenous resources (Sec.~\ref{sec:disaggregation}).
%
% WE DO NOT DO THAT!
%We show that the partitioning of shared resources helps isolate the interference,
%diminishing co-location performance losses and providing a simple yet fair cost model.}
%
Our system allocates functions on idle resources while requiring changes to neither the
hardware nor the operating systems.
Using functions executing in isolated containers, we can securely share node resources between users.
%
%we propose a new variant of a \textbf{\emph{job stripping}} where batch workloads are co-located
%with FaaS functions to boost system utilization (Sec.~\ref{sec:disaggregation}).
%
We then define the requirements that HPC functions must fulfill to overcome
the limitations of the classical, cloud-oriented functions.
We present an \textbf{HPC-centric FaaS platform} by adapting a high-performance serverless runtime rFaaS~\cite{copik2021rfaas} to Cray supercomputers and HPC containers, and enhance it with a performance-oriented \textbf{programming model and integration} (Sec.~\ref{sec:faas}).
%
%and HPC-centric \textbf{programming model and integration} (Sec.~\ref{sec:faas}).
%
%Finally, we present an HPC-centric \textbf{programming model and integration} for FaaS (Sec.~\ref{sec:programming_model}).
%
We use the pools of idle memory to host function containers,
reducing cold startups and increasing resource availability.
We evaluate the new system on representative HPC benchmarks (Sec.~\ref{sec:evaluation}).
To the best of our knowledge, our work is the first FaaS solution specialized for HPC environments and evolving and malleable jobs.

% our solution
Our paper makes the following contributions:
\begin{itemize}
  \item We introduce a novel co-location strategy for HPC workloads that improves system utilization and uses pools of underutilized memory to host function containers.
  \item We adapt a high-performance FaaS platform to supercomputers
    %and show that HPC-native function invocations scale to thousands of cores.
    and demonstrate the efficiency of HPC functions.
  \item We present an integration of FaaS into the HPC batch scheduling system and the MPI programming model, and show how functions accelerate HPC applications.
\end{itemize}

\section{Background and Motivation}

Serverless provides a new resource allocation paradigm that can mitigate the low resource
utilization (Sec.~\ref{sec:background_utilization_hpc}).
Functions provide a software approach to fine-grained resource allocations, overcoming the disadvantages of hardware solutions (Sec.~\ref{sec:background_resource_disaggregation}).
%
%Simultaenously,
Functions can improve on the existing techniques and billing systems
for co-locating workloads (Sec.~\ref{sec:background_colocation}).

\subsection{Resource Utilization in HPC}
\label{sec:background_utilization_hpc}
To assess the modern scale of the HPC underutilization,
we analyzed the Piz Daint supercomputer, and disentangled
the CPU and memory utilization in Figures~\ref{fig:daint_cpu} and~\ref{fig:daint_memory}, respectively.
While the median number of idle nodes at any sampling point was 252,
the rapid and frequent changes indicate that resources do not stay idle long.
The median availability time is between 5 and 6.5 minutes, and 70-80\% of unallocated nodes stay idle for less than 10 minutes (Fig.~\ref{fig:daint_idle}); similar results were observed on other systems~\cite{10046086}.
\textbf{This gap cannot be addressed with persistent and long-running allocations.}

The aggregated and statically allocated computing nodes lead to wasting
memory and network resources~\cite{9235066,10.1145/3352460.3358267,10.1145/3514245}.
The average node memory usage can be as little as 24\%, while the memory system contributes roughly 10-18\% of the appropriation and operational expenditures~\cite{10.1145/3023362,barroso2018datacenter}.
Furthermore, network and memory bandwidth utilization is very low, with occasional
bursts of intensive traffic~\cite{10.1145/3514245}.
Unfortunately, this problem is fundamentally not solvable with current static allocations
on homogeneous resources because these do not represent the heterogeneity of HPC workloads.
While capacity computing applications with poor scaling require gigabytes of memory per process,
capability computing can use less than 10\% of node memory~\cite{10.1145/3023362}.
The differences between MPI ranks and applications add further imbalance.

Heterogeneity of HPC systems is increasing over time~\cite{10.1145/3432261.3432263},
with five more TOP500 systems using GPUs every year~\cite{top500}.
However, actual GPU utilization is often quite low.
For example, on the Titan system, only 20\% of the overall jobs used GPUs~\cite{8891001}.
Furthermore, some applications that use GPUs make no use of the CPUs,
reinforcing a need to co-locate GPU and CPU workloads~\cite{10.1145/3432261.3432263}.

\begin{bluebox}
HPC resources are underutilized and overprovisioned due to diversity of workloads.
Batch jobs cannot use idle computing resources due to their short availability.
%and the diverse workloads force node overprovisioning.
\end{bluebox}

\subsection{Resource Disaggregation}
\label{sec:background_resource_disaggregation}
Remote and disaggregated memory has been considered in data centers for
almost a decade now~\cite{10.1145/1555754.1555789,179767,201565,199305,10.1145/3127479.3131612}.
Disaggregation replaces overprovisioning for the worst case with allocating for the
average consumption but retaining the ability to expand resources dynamically.
Remote memory has been proposed for HPC systems~\cite{9235066},
but it comes with a bandwidth and latency penalty.
While modern high-speed networks allow retaining near-native performance in some applications~\cite{199305},
remote memory is considered challenging for fault tolerance, and performance reasons~\cite{10.1145/3127479.3131612}.

Hardware-level solutions can elevate performance issues, e.g., by providing
a dedicated high-speed network~\cite{9252003}
and using dedicated memory blades~\cite{10.1145/1555754.1555789}.
However, many methods have not been adopted because of the major investments needed~\cite{201565},
such as changes in the OS and hypervisor, explicit memory management, or hardware support
~\cite{4154093,10.1145/1555754.1555789,10.1145/1555754.1555789,9235066}.

\begin{bluebox}
Resource disaggregation is not common in HPC because of performance
overheads and increased complexity.
\end{bluebox}

\subsection{HPC Co-location}
\label{sec:background_colocation}
Co-location mitigates the underutilization problem by allowing more than one batch job
to run on the same node.
While some studies have not found a significant difference between node-sharing
and exclusive jobs~\cite{10.1145/2616498.2616533,10.1145/2949550.2949553},
many applications experience performance degradation through contention in
shared memory and network resources~\cite{10.5555/2388996.2389109,6468475}.
Prior work has attempted to improve scheduling on a node by detecting sharing and contention in shared interfaces
%the memory bus, bandwidth, and network interface
~\cite{10.1145/511334.511343,10.1145/566726.566768,1655680,10.1145/1854273.1854306,1240622}.

\emph{Symbiotic applications} can improve their performance when co-located~\cite{10.1007/978-3-540-71035-6_10,10.1145/511334.511343,https://doi.org/10.1002/cpe.3187},
but determining which workload pairs show positive symbiosis is hard.
Methods include user hints and offline experiments~\cite{10.1007/978-3-540-71035-6_10,10.1145/1183401.1183450},
profiling and online monitoring~\cite{1655680,1240622,10.1145/2499368.2451126},
and machine learning~\cite{10.5555/2388996.2389109}.
For co-location, systems should select applications with different characteristics~\cite{5289162,10.1007/978-3-540-71035-6_10,10.1145/1183401.1183450}.
\emph{Job striping}
provides further performance benefits
by spreading application processes and co-locating them with other workloads~\cite{https://doi.org/10.1002/cpe.3187,5289162}.
Another difficulty imposed by sharing is the unfairness of traditional billing
models when applied to jobs with performance impacted by the interference~\cite{https://doi.org/10.1002/cpe.3187,6877470}.
Finally, sharing introduces security vulnerabilities when tenants are not isolated.

\begin{bluebox}
  Node sharing is beneficial for the efficiency of HPC,
  as long as it avoids harmful interference.
  Short functions are good candidates for interference-aware co-location.
\end{bluebox}

\section{Software Disaggregation with FaaS}

\label{sec:disaggregation}

\begin{figure}[!t]
	\centering
  \includegraphics[width=1.0\linewidth]{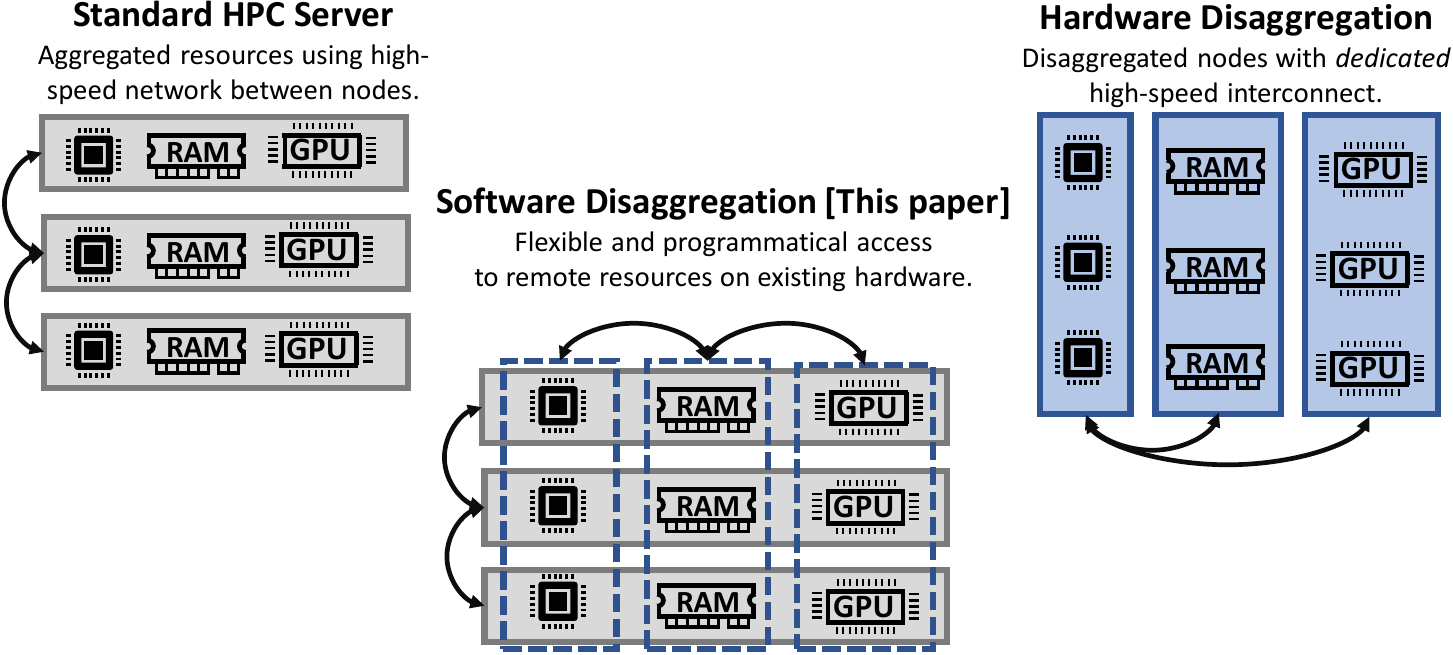}
  \caption{\textbf{Software disaggregation}: co-location provides 
    semantics of resource disaggregation on an unmodified system.}
  \label{fig:disagg_comp}
\end{figure}

Software disaggregation allows jobs to access remote resources by invoking serverless functions (Fig.~\ref{fig:disagg_comp}).
This new approach is flexible and targets only underutilized nodes.
In \emph{hardware} disaggregation, HPC applications always pay the latency penalty of accessing remote resources.
In \emph{software} disaggregation, standard HPC applications still run on unmodified nodes and have all hardware resources available locally.

We begin by disaggregating resources available on \textbf{idle} nodes.
Then, we go further and handle unused resources within active nodes.
We focus on the three resources that can be disaggregated: \textbf{CPU cores},
\textbf{memory},
and \textbf{GPUs}.
As in \emph{job striping}, where users spread processes across a larger number of nodes to increase throughput, we recommend the same approach to leave at least one core free on each node to access idle memory and GPUs without introducing temporary oversubscription.
We co-locate long-running jobs with short-term, flexible tasks with intensive but complementary resource consumption,
Serverless functions are perfect for co-location: they offer fine-grained scaling,
multi-tenant isolation and are very easy to checkpoint, snapshot, and migrate.

\subsection {Reusing Idle Nodes}
\label{sec:colocation_idle}
\vspace{-0.75em}
\begin{figure}[h!]
\includegraphics[width=0.95\linewidth]{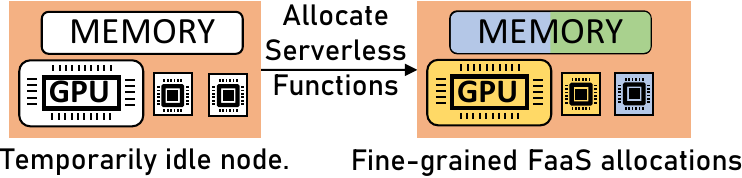}
\end{figure}
\vspace{-0.75em}
While shortly available nodes are impractical for batch jobs,
they can be utilized for time-limited functions that often require just a few seconds to execute~\cite{254430}.
Once a node is available, we deploy there a batch job with a serverless worker to start accepting function invocations.
A single multi-core node can support concurrently many fine-grained functions that target newly available CPU cores, memory, and GPUs.

This scenario puts three requirements on the serverless platform: it has to integrate a new node
quickly, release it immediately when the batch system needs it, and gracefully handle the node termination.
We use the high-performance serverless platform rFaaS that uses leases to manage ephemeral allocations (Sec.~\ref{sec:faas}).
Once the node has to be returned, a signal sent to the rFaaS executor blocks any new invocations and waits for current, time-limited functions to finish.
At the same time, the executor cancels existing leases, notifying the client libraries to redirect further requests to a new lease.

\subsection {Co-location - Sharing CPUs and more}
\label{sec:colocation_cpus}
\vspace{-0.25em}
\begin{figure}[h!]
\includegraphics[width=0.95\linewidth]{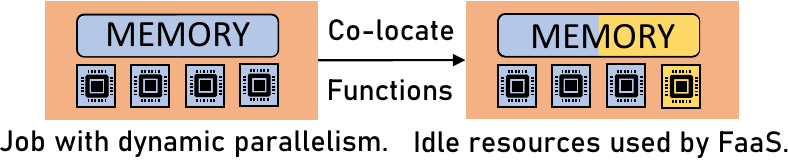}
\end{figure}
\vspace{-0.75em}
Scientific applications often have constraints on the parallelism and problem size
%or the problem size
beyond those imposed by the hardware.
For example, LULESH~\cite{osti_1117905} must use a cubic number of parallel processes,
making job configurations unlikely to perfectly match the available number of cores.
Furthermore, the co-location of many MPI ranks leads to contention~\cite{10.1145/3437801.3441613}, forcing users to spread processes across nodes.

We improve utilization by locating FaaS executors to target unused cores in a node. % (Fig.~\ref{fig:disaggregation_1}).
Thus, our new serverless approach implements \emph{job striping}, where MPI processes
do not occupy an entire node and are co-located with other applications to better utilize resources~\cite{https://doi.org/10.1002/cpe.3187,5289162}.
Functions can use the rest of the node's resources while minimizing the performance impact on
the batch application.
Since FaaS functions are easy to profile and characterize, they can be matched with jobs that
present different resource availability patterns.
Even when resource consumption cannot be aligned, partitioning shared memory and CPU resources can provide fairness
%needed
for each application.

Furthermore, short-running MPI processes resemble FaaS functions (Sec.~\ref{sec:programming_model_mpi}).
Adaptive MPI implementations~\cite{10.1145/3075564.3075585,10.1145/2966884.2966917}
rescale applications by adding and removing processes on the fly,
and new MPI ranks can be allocated in a serverless fashion.
We demonstrate the benefits of co-locating such MPI processes with the example of the NAS benchmarks (Sec.~\ref{sec:evaluation_colocation}).

\subsection{Memory Service for Applications}
\label{sec:colocation_memory}
\vspace{-1em}
\begin{figure}[h!]
\includegraphics[width=0.95\linewidth]{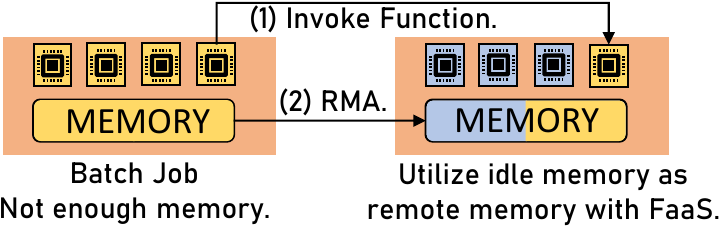}
\end{figure}
\vspace{-0.75em}
In HPC, the memory usage of a job varies between processes and
within the lifetime of the job, with a difference of up to 62.5x for some applications~\cite{10.1145/3023362}.
Furthermore, applications with poor scaling require gigabytes of memory per process,
while capability computing can use less than 10\% of available memory~\cite{10.1145/3023362}.
Therefore, HPC nodes will always have overprovisioned memory to support
heterogeneous workloads.
While high-memory jobs are not frequent in HPC systems, they still need to be accommodated,
requiring idle memory reclamation to be short-term and ephemeral.

We propose three methods to effectively use idle node memory in HPC applications.
First, we use free memory to keep FaaS containers warm and allow functions
to be started quickly and efficiently, resolving an important issue of expensive
cold starts in serverless (Sec.~\ref{sec:faas_cold_starts}).
Then, we use the memory to host object storage nodes (Sec.~\ref{sec:faas_context}).
Finally, we offer other jobs the ability to run \textbf{memory service
functions}. % (Fig.~\ref{fig:disaggregation_2}).
Functions allocate a memory block and offer direct access, allowing HPC applications for remote paging~\cite{9235066}.
We use one-sided remote memory access (RMA), which adds minimal CPU overhead to the system~\cite{201565}.
Thus, many memory service functions can be co-located, even with compute-intensive applications.
%such as LULESH.
%
%Modern networks provide RMA with low latencies~\cite{201565},
Functions enable memory service with fine-grained scalability, easily
controllable lifetime, and multi-tenant isolation.

When the batch system needs to reclaim idle memory, function containers
can be migrated to other nodes and swapped to the parallel filesystem.
The client library can make submitting functions seamless for the user,
with functions running either directly from warm containers in otherwise idle memory
or loaded from the swapped container if necessary.

\subsection{GPU Sharing}
\label{sec:colocation_gpus}
\vspace{-1em}
\begin{figure}[h!]
\includegraphics[width=0.95\linewidth]{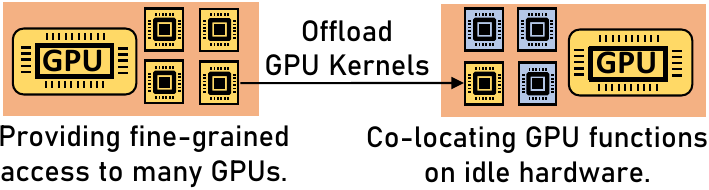}
\end{figure}
\vspace{-0.75em}
While the heterogeneity of HPC systems is growing, not every application can
be modified to benefit from GPU acceleration.
HPC systems should co-locate CPU-only and GPU-enabled jobs,
as these are often complementary~\cite{10.1145/3432261.3432263}.
For example, the main version of LULESH does not use accelerators at all,
instead relies on MPI and OpenMP.

We disaggregate GPU and CPU resources by co-locating GPU functions.
The function can be co-located with CPU-only functions and applications, requiring only a single CPU core to manage device and data transfers.
Such functions can be launched with containers specialized for HPC systems (Sec.~\ref{sec:faas_containers}).
Furthermore, functions can keep warm data in the device's memory until
another application needs the device.

Although there exist systems for remote GPU access~\cite{duato2010rcuda},
they add latency to each command.
However, applications such as machine learning inference can consist of hundreds
of kernels with synchronization in between~\cite{toblerthesis}.
By running one CPU function process to ensure GPU access, we avoid adding inter-kernel
latency in the remote GPU scenario.

\begin{figure}[t]
  \centering
  \includegraphics[width=\linewidth]{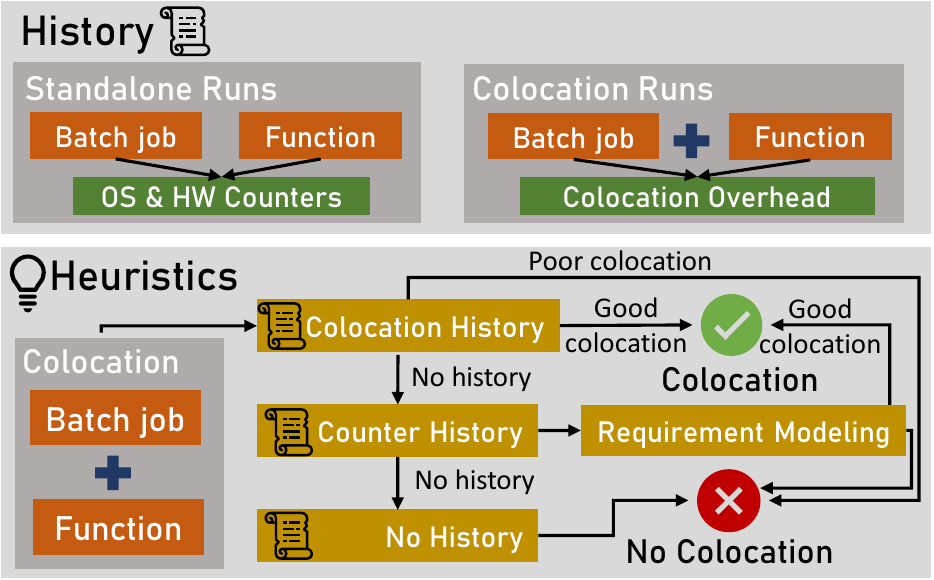}
  \caption{Co-location policies use lightweight online monitoring.}
  \label{fig:colocation_heuristic}
\end{figure}

\subsection{Co-location Policies}
To decide if a function can be co-located with a given job, we must consider two factors: availability of shareable resources on a node and potential interference.
For the latter, we propose
%identifying the most stressed resource classes and 
selecting applications that do not stress the same resource simultaneously.
To that end, we introduce a design based on the practical job history and a heuristic using the well-studied methods of HPC performance modeling (Fig.~\ref{fig:colocation_heuristic}).

\noindent\textbf{Availability.}
Disaggregation is an entirely \emph{opt-in} system policy where users voluntarily share the node to obtain lower computing prices.
We do not modify SLURM but use existing features: to enable co-location, users set the SLURM \code{shared} flag or submit the job to a designated partition.
Job core count and memory size determine the allocated resources on each node, and serverless functions can use the remaining ones.
We use the SLURM GRES to determine how many and which GPUs are available~\cite{slurmGRES}.
We do not consider GPU sharing due to security and interference issues~\cite{10.1145/3433210.3453077,GILMAN2021102234}.
Instead, GPU virtualization and partitioning can create isolated sub-devices in the GRES system.

\noindent\textbf{History.}
First, we establish which workload pairs could be co-located.
Estimating performance interference between any two applications is a complex problem~\cite{10.1145/2556583}.
Fortunately, the problem can be simplified because HPC systems serve a limited number of applications, e.g., about 115 (Blue Waters)~\cite{DBLP:journals/corr/JonesWIDSPGFSBK17} and 650 (NERSC Hopper)~\cite{zhao2013effects,antypas2013nersc}.
Furthermore, HPC applications are invoked many times with varying parameters, often by multiple tenants.
We can cover two-thirds of the total computation time by analyzing no more than 25 applications~\cite{DBLP:journals/corr/JonesWIDSPGFSBK17,antypas2013nersc}.

Thanks to the limited workload diversity, we can keep a global history available to the serverless resource manager.
For each co-location, we record the runtime of the batch job and the function, and compare it later against an exclusive run with the same parameters.
A lightweight sampling of hardware and operating system counters gathers information on the FLOPs, memory accesses, and network traffic.
Thanks to the shorter runtime and encapsulated form, serverless functions can be executed independently.
Thus, when registering a new code container, the function can be profiled using user-provided or synthetic input data~\cite{10.1145/3588195.3592996}.
Operators can make function profiling mandatory or compensate users for the additional work of supplying information.
Finally, counter data can be used to detect poor utilization and recommend colocation to users. %users to colocate jobs with poor utilization.

\textbf{Heuristics}
We use the colocation history as a primary metric for estimating interference overhead.
When the history is unavailable for the first colocation instance, we apply \emph{resource} \emph{requirement} \emph{modeling}~\cite{8514881}.
This method uses counter measurements to create performance models for different resource classes, allowing us to compare the stress factors for each application.
Since models are created in the background, we can remove modeling from the critical path of scheduling.
Subsequent serverless invocations will be decided with the help of history entries generated from the first co-located run.
In our approach, users should be incentivized with lower computation prices to provide additional information, such as specifying their application and inputs.
Given the limited number of distinct applications, we can practically support many co-location combinations.
Furthermore, disaggregation can be composed with methods for estimating performance interference~\cite{10.1007/978-3-319-96983-1_4} and dedicated pricing models for co-location~\cite{6877470}.

\subsection{Interference and Sharing Fairness}
Resource sharing can introduce performance interference,
a concern for large-scale jobs that are significantly affected by network and OS noise~\cite{hoefler2010characterizing,5161095,10.1145/3570609}.
However, the guarantee of exclusive access to
%on-node
resources is illusive, as jobs
are affected by the inter-node sharing of network resources~\cite{6877474,7877142}.
The disaggregation must consider only the contention on node resources,
as the user cannot control network performance.

We propose that the disaggregation is applied selectively to workloads,
and \emph{hero} jobs are exempted since they allocate a large fraction of the entire system and can be sensitive to interference.
Since many jobs use less than 256 nodes~\cite{DBLP:journals/corr/JonesWIDSPGFSBK17,10.5555/3433701.3433812},
disaggregation can target small and medium-scale jobs, increasing system throughput while not impacting scalability.

%\section{HPC FaaS Runtime}
\section{HPC FaaS}

\label{sec:faas}

\begin{figure}[t]
	\centering
  \includegraphics[width=1.0\linewidth]{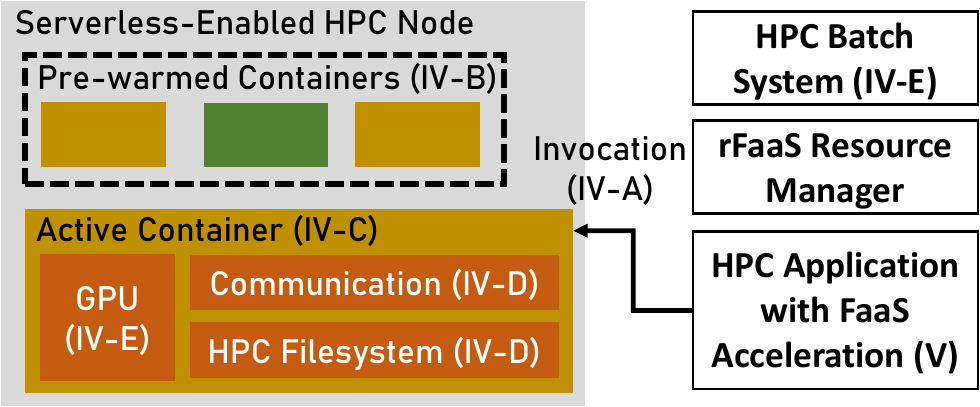}
  \caption{Specializing serverless platform for HPC requirements.}
  \label{fig:hpc_faas}
\end{figure}

Serverless computing brings an abstract view of data center resources
allocated on the fly by the provider and hidden from the user.
This abstraction frees users from any responsibility for provisioning
and allows for elastic computing, where users are billed only for the resources used.
FaaS is the dominating programming model where users invoke stateless functions to the cloud.
However, \emph{classical cloud functions} have been designed for the hardware and software
stack common in the cloud.
The situation changes in supercomputing systems with performance-oriented
architecture and programming models.
We map cloud functions into HPC environments and identify five major issues that serverless faces in high-performance systems  (Table~\ref{tab:cloud_hpc_functions}).
Based on these results, we define requirements that \emph{HPC functions} must meet
to become an efficient component of a high-performance application (Fig.~\ref{fig:hpc_faas}).

\textbf{rFaaS}.
To demonstrate how serverless functions can be used in the HPC context, we extend the serverless platform rFaaS~\cite{copik2021rfaas}.
rFaaS allows consecutive invocations to execute on the same resource allocated with a temporary lease.
Furthermore, it employs a direct RDMA connection between the client and function executor, optimizing both the latency and the bandwidth of serverless.
%
%However, building a portable function environment on a supercomputing system is technically challenging, mainly due to changing software environments and the restricted execution model of batch jobs, designed primarily for static and long-running applications.
However, building a portable function environment on a supercomputing system is technically challenging, mainly due to specialized hardware and restricted execution models designed for static and long-running applications.
We demonstrate that a serverless platform can be used efficiently in an environment that defaults towards exclusive jobs with implicit resource assignments and homogeneous applications.
To that end, we present a specialization of the rFaaS platform to the Cray XC40/XC50 system Piz Daint~\cite{pizDaint}.

\begin{table}
  \footnotesize
  \centering
  \begin{adjustbox}{width=\linewidth}
  \begin{tabular}{lccc}
    \toprule
                    & Cloud FaaS   & HPC FaaS      \\
    \midrule
    Network         &   TCP     & \textbf{uGNI}, ibverbs, AWS EFA\\
    Sandbox         & Docker, microVM   & Singularity, \textbf{Sarus}\\
    Storage         & Object, block         & Parallel file system\\
    Communication   & Storage, DB, queue    & Direct communication\\
    Placement       &  VMs, Kubernetes   & Batch jobs on HPC nodes\\
    \bottomrule
  \end{tabular}
  \end{adjustbox}
  \caption{Comparison of \emph{cloud functions} environments with \emph{HPC functions}. Technologies used in specialization for Cray machines are in bold.}
  \label{tab:cloud_hpc_functions}
\end{table}

\subsection{Slow Warm Invocations}
\label{sec:faas_interconnect}
\noindent\textbf{Problem}
Each invocation of a classical function requires centralized scheduling and rerouting payload to a selected function sandbox.
%When an invocation of a classical function is \emph{triggered},
%the payload is redirected to a selected function sandbox.
%
Even a \emph{warm invocation} in an existing sandbox can introduce
dozens of milliseconds latency~\cite{copik2021sebs}.
%due to centralized rerouting and lack of high-speed network transport.
%~\cite{copik2021rfaas}.
%
However, functions require microsecond-scale latency to benefit from computing on remote resources (Sec.~\ref{sec:integration}).

\noindent\textbf{Solution}
\toolname{} uses fast networks and a shortened invocation critical path to achieve single-digit microsecond latencies.
%
%Similarly to MPI implementations, the serverless platform needs to use the
%features the interconnect has available to offer competitive performance.
%
To deploy rFaaS on a Cray system, we use the \emph{libfabric} to target \emph{uGNI}, the user network interface for Cray interconnects~\cite{pritchard2016gni}.
We faced two major challenges: first, the \emph{libfabric} installation within a container must be replaced with the main system installation to achieve high performance and manage access to uGNI.
%
% https://www.sciencedirect.com/science/article/pii/S0743731522000387?casa_token=YYjHwt2xixIAAAAA:exOjIRLn5koqvQOg4xIycpoUWdFzP7hKqtPJcfoYMvJwHR_EggrrtSMBz0vceAh3_DS7vFkR-g
We resolve the issue by manually mounting system directories in the container, as the available HPC containers do not support injection of \emph{libfabric} at the runtime~\cite{10029965}.
Furthermore, \emph{uGNI} is designed to communicate within a single batch job,
which is not the case in FaaS: the client in one job communicates with an executor running in another batch job.
To support functions on the Cray system, we implement the allocation and distribution of security credentials DRC~\cite{shimek2016dynamic}.
%
%~\cite{HUANG2022106}

\subsection{Expensive Cold Starts}
\label{sec:faas_cold_starts}
\noindent\textbf{Problem}
When no existing sandbox can handle an invocation, a new one is allocated and initialized with user code.
%with an executor process running the user code.
%
This \emph{cold start} has a devastating effect on performance since it adds
hundreds of milliseconds to the execution time in the best
case~\cite{coldStart,10.1145/3423211.3425682,copik2021sebs}.
Standard mitigation techniques include lightweight and prewarmed sandboxes
and faster bootup methods~\cite{216031,246288,10.1145/3373376.3378512},
but the most common one is retaining containers for future invocations.
However, its effectiveness is limited as idle containers are purged to free memory.
%ccupy memory and
%are purged frequently.

\noindent\textbf{Solution}
Instead of decreasing negative cold start effects, we focus on reducing their frequency with the
the help of unutilized node memory.
%
%In HPC systems, we can benefit from ubiquitous memory overprovisioning
%to keep containers alive for a long time.
%
This solution is compatible with batch systems and fits the short-term availability of resources perfectly
because idle containers can be removed immediately without consequences.
The availability of CPU cores to handle invocations can be guaranteed by modifying allocations
to keep one or two cores per node (out of the 30 or more) available.
%
%Furthermore, functions can be scheduled on busy nodes with oversubscription,
%which usually increases system throughput~\cite{5470434}.
%
%We modify the allocation algorithm for rFaaS.
%
%First, the resource management system is extended with information on available
%containers and their expected lifetime, and decentralized scheduling takes
%this into account.
%We modify the rFaaS resource management to remember the retained containers
%and adjust the allocation algorithm to target nodes with warm containers.
We adjust the rFaaS resource management to target nodes with warm containers.
%
%Then, the cold start overhead is dominated by establishing RDMA connection and not
%by the expensive initialization of a new container.
Then, the cold start overhead is dominated only by establishing RDMA connection.
%and not
%by the expensive initialization of a new container.

\begin{table}[t]
  \footnotesize
  \centering
  \begin{tabular}{lccc}
    \toprule
                              & Docker      & Singularity       & Sarus\\
    \midrule
    Image Format              & Docker            & Custom      & Docker-compatible\\
    Repositories              & Docker registry   & None        & Docker registry \\
    Devices support           & Through plugins   & Automatic   & Automatic\\
    Resources                 & Native, cgroups    & Automatic   & Automatic \\
    Batch System              & None              & Slurm       & Slurm\\
    MPI Support               & None              & Native      & Native    \\
    \bottomrule
  \end{tabular}
  \caption{Comparison of container systems for cloud and HPC~\cite{Kurtzer2017,benedicic2019sarus}. Automatic resource and device support in Singularity and Sarus are done via Slurm.}
  \label{tab:container_systems}
\end{table}

\subsection{Incompatible Container Systems}
\label{sec:faas_containers}
\noindent\textbf{Problem}
Serverless in the cloud is dominated by Docker containers and virtual
machines~\cite{246288}.
However, the adoption of containers is limited by security concerns,
and virtual machines limit access to the accelerator and network devices.
%necessary for high-performance applications.
%
Containers must run in the \emph{rootless} mode to avoid
privilege escalation attacks.
To support multi-tenancy in HPC, these issues must be mitigated while
retaining near-native performance.

\noindent\textbf{Solution}
Serverless sandboxes must be tailored to the needs of HPC functions,
and we consider containers designed for scientific computing:
Singularity~\cite{Kurtzer2017} and Sarus~\cite{benedicic2019sarus}.
Both provide native access to compute and I/O devices
and integrate batch resource management (Table~\ref{tab:container_systems}).
Furthermore, containers provide native support for high-performance MPI
installations with dynamic relinking.
%of applications.
%
This enhancement is essential for HPC functions to support elastic execution
of MPI processes.

\subsection{Lack of a High-Performance I/O}
\label{sec:faas_context}
\noindent\textbf{Problem}
Classical serverless functions cannot accept incoming network connections in the cloud
as they operate behind the NAT gateway and have no public IP address.
Instead, functions must resort to using persistent cloud storage, with latencies in the tens of
milliseconds, and transmitting results back to the invoker --- no high-performance
I/O is available to the functions in the data center ecosystem.
However, HPC applications can produce terabytes of data, and in such applications,
the transmission of results from a function to the invoking MPI process
quickly becomes impractical.
In HPC, high-performance I/O is offered through the scalable parallel filesystem~\cite{10.1145/3309205}, replacing the need for cloud storage.
Furthermore, this environment is too restricted for memory service functions
that accept incoming RDMA connections.
%Furthermore, this environment is too restricted for remote memory functions
%that accept incoming connections and return the memory buffer information
%to the user without ending the invocation.

\noindent\textbf{Solution}
First, we make the parallel filesystem accessible by mounting the user's partitions in the function container. %and allowing functions to access the user's data.
This allows for storing large amounts of data from invocations and brings serverless performance in line with what is expected of HPC applications.
We continue to use object storage as a warm cache for lower latency on small files (Sec.~\ref{sec:evaluation_rfaas}).
%co-locating storage on nodes with idle memory.
%Object storage nodes can be placed on nodes with idle memory, providing warm caches for lower latency on small files (Sec.~\ref{sec:evaluation_rfaas}).
%
Then, we enhance functions with a portable interface to start communication, accept incoming connections, and return data without terminating the invocation,
letting HPC users implement functionalities that do not fit the classical cloud model, such as remote memory.% service.
%
%Furthermore, we enable returning data from functions that continue the invocation,
%allowing HPC users to implement functions that do not fit the classical cloud model.

\subsection{Incompatible Resource Management}
\label{sec:faas_batch_integration}
\noindent\textbf{Problem} Serverless platforms allow to configure memory size, with CPU resources allocated proportionally to memory~\cite{awsAPIBiling,gcpPricing}.
However, software disaggregation techniques require allocations of one hardware resource
while not using another one extensively.
Using cluster resources requires two new functionalities: a release of nodes for FaaS processing and the removal of executors from the serverless resource pool.

\begin{figure}[t]
	\centering
  \includegraphics[width=1.0\linewidth]{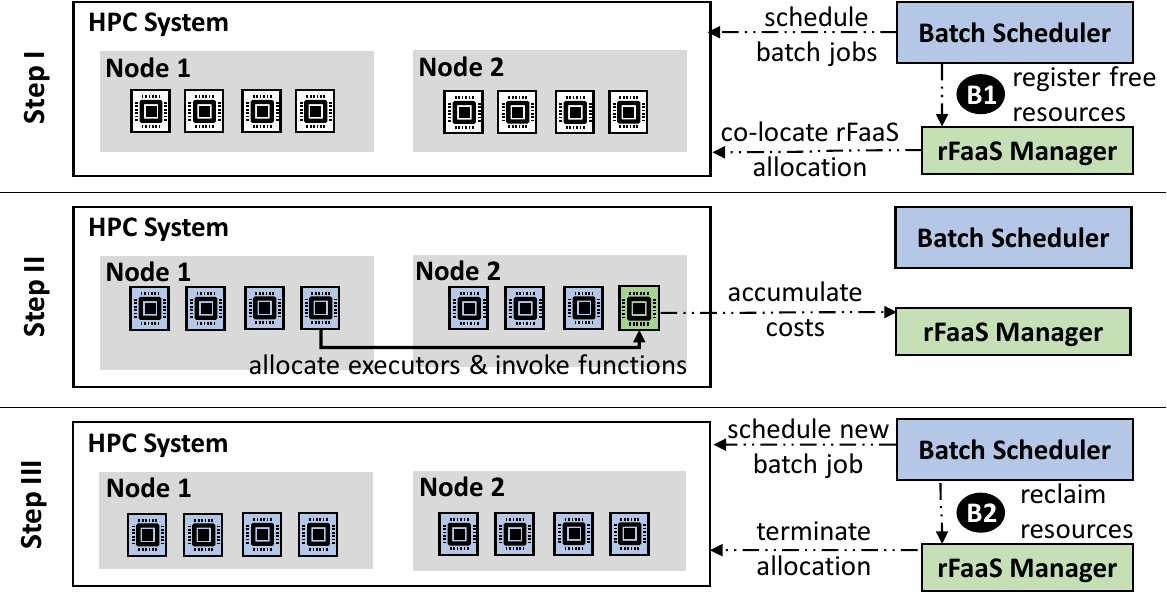}
  \caption{Co-location made easy: \toolname{} functions
  running on batch-managed clusters.}
  \label{fig:system_model}
\end{figure}

\noindent\textbf{Solution}
We extend the rFaaS resource management with memory and GPU device
availability.
Computing and memory resources are allocated and billed independently:
users configure memory size according to needs and add a GPU device.
Since we operate on reclaimed idle resources, there is no monetary loss from partial resource consumption by functions: every allocation increases system utilization.

Then, we implement an interface in \toolname{} designed
for integration with cluster job management systems (Fig.~\ref{fig:system_model}).
The global resource manager offers a single REST API call to register
resources (\circledColorSmall{B1}{black}), which are used immediately,
supporting allocations on the spare capacity available only for a very short time (Fig.~\ref{fig:daint_stats}).
Released resources include CPU cores, memory, and GPUs that have not been
explicitly allocated by the tenant.
The allocation policy becomes \emph{opt-in} - resources not
requested by the user are not assigned by default to their jobs.

Furthermore, we allow the batch manager to retrieve resources for jobs with higher priority.
A REST API \emph{remove} call starts resource deallocation (\circledColorSmall{B2}{black}).
%and it has a parameter describing
%the allowed termination time 
%Each \toolname executor implements a REST API \emph{remove} call with a single parameter
%describing the allowed time for resource deallocation (\circled{B2}).
%
When the request is immediate (no additional computing time is allowed), all active function invocations are aborted, and \emph{termination} replies are sent to clients.
%through existing RDMA channels,
%and the final billing update is sent to the resource manager.
%during shutdown of the execution sandbox.
%
Otherwise, active invocations are allowed to finish, but no further invocations will be granted.

\subsection{Incompatible Programming Model}
\label{sec:integration}
\noindent\textbf{Problem}
Functions are designed for event-based programming in the cloud,
and we need a new performance-oriented approach to benefit from serverless in HPC.
While applications can use fine-grained invocations to offload computations,
the performance depends on the cost of moving data and waiting for a remote task.
Thus, we need a model to tell us \emph{when} remote invocations can be integrated into
HPC applications.

\noindent\textbf{Solution}
We propose an integration of \toolname{} invocations based on
\emph{LogP} models~\cite{10.1145/155332.155333,hoefler-pmeo-06}.
The guiding principle -- the application never waits for remote invocations to
finish -- is achieved by dividing the work such that the network transport and computation times are hidden by local work.
%
%The network performance is expressed with parameters such as latency,
%CPU overhead on the sender and receiver, and gap factor. %determining the minimum time between messages.
%
We learn the network parameters, estimate the compute time of offloaded tasks,
and measure the \toolname{} overheads.
%we model the round-trip invocation time.
%
%We design a model to decide \emph{when} remote invocations can be integrated into
%HPC applications, and then show \emph{how} to use \toolname{} as an accelerator for HPC.
%
%The model is applied to each offloaded task separately to support the varying
%computational and I/O requirements of heterogeneous applications.
%
We provide a non-exhaustive list of examples that %are either natural
%fit or
can be adapted to offloading.
%We provide examples of applications that are either natural
%fit or can be adapted to serverless offloading.
%
%This non-exhaustive list provides an intuition on using serverless efficiently in HPC.
%}

\setlength{\abovedisplayskip}{3pt}
\setlength{\belowdisplayskip}{3pt}

\emph{Massively parallel applications}
These applications are extremely malleable and can efficiently offload tasks as functions.
A solver for the Black-Scholes equation~\cite{10.1080/00207160.2012.690865} is a good example,
as it generates many independent tasks with comparable runtime.
To achieve the best possible performance, we measure the runtime of one
task $T_{local}$ and then compare this to the runtime $T_{inv}$ of one invocation
using \toolname{}, to which we add the round-trip network time $L$. %of sending the task and receiving the result $L$.
Time $T_{local}$ can be obtained with offline profiling tools common
in performance modeling~\cite{10.1145/3437801.3441613,10.1145/2503210.2503277},
providing measurements and models for runtime decisions without the overhead
of additional invocations.
There exists a number $N_{local}$ of tasks such that:
\begin{equation}
N_{local} \cdot T_{local} \ge T_{inv}  + L
\label{eq:offload}
\end{equation}
Therefore, if the number of tasks is greater than $N_{local}$, up to $N_{remote}$ tasks can
be computed remotely without incurring any waiting time.
$N_{remote}$ is determined as the number of tasks necessary to saturate the available bandwidth $B$:
$\frac{B}{Data_{inv}}$.
Therefore, the throughput of the system only depends on the network link bandwidth
and the amount of work available.

\emph{Task-based applications with no sharing}
%
%Task dependency graph~\cite{10.1145/3018743.3018770} specifies the order of execution
%and dependencies between tasks in a program, which can be offloaded using the guideline %in Eq.~\ref{eq:offload}.
Task dependency graph~\cite{10.1145/3018743.3018770} specifies the order of execution of tasks, which can be offloaded using Eq.~\ref{eq:offload}.
%
%However, the number of tasks that can be offloaded depends on the width of the task
%dependency graph - the wider the graph, the more parallelism is exposed,
%and therefore, more tasks can be transferred to \toolname{}.
However, the number of tasks that can be offloaded depends on the width of the task
dependency graph - the wider the graph, the more parallelism is exposed.
%and therefore, more tasks can be transferred to \toolname{}.
%
As an example, we consider the distributed prefix scan in electron microscopy
image registration~\cite{copik2020workstealing},
where the width of the task graph varies significantly between program phases.
%
%Thus, serverless can achieve higher efficiency than static allocation.
%
%The width of the task graph in a distributed scan varies significantly between program phases; thus, dynamic serverless offloading achieves higher efficiency than static resource allocation.

%Second, we propose to run MPI applications \emph{as functions},
%providing a backend for short-running computations on idle hardware (Sec.~\ref{sec:programming_model_mpi}).

\emph{MPI Functions}
\label{sec:programming_model_mpi}
An HPC function can also implement the same computation and communication logic as an MPI process.
These can be allocated with lower provisioning latency than through a batch system,
and use computing resources with short-time availability.
New MPI ranks can be scheduled as functions without going through the batch system,
implementing the infrastructure needed to support adaptive MPI~\cite{10.1145/2966884.2966917,9406729}.
In HPC, FaaS can be more than a backend for website and database functionalities; functions can represent full-fledged computations with communication and synchronization~\cite{10.1145/3577193.3593718}. %bringing low bootup times and flexible resource management to \emph{evolving} and \emph{malleable} jobs~\cite{10.1007/BFb0022284}.

%When running in a sandbox, serverless functions can execute on a multi-tenant node,
%resolving one of the major security concerns that prevent node sharing in a production system.
%
%In HPC, FaaS can be more than a backend for website and database functionalities; functions can represent full-fledged computations with communication and synchronization.
%
%Serverless brings low bootup times and flexible resource management to \emph{evolving} and \emph{malleable} jobs~\cite{10.1007/BFb0022284}, desired traits to scale up in the application phase with adaptive parallelism.

%A further benefit can be provided with support for adaptive MPI implementations.
%
%These usually require infrastructure extensions to support elastic scaling~\cite{10.1145/2966884.2966917,9406729}.
%
%Instead, new MPI ranks can be scheduled as functions without going through the batch system.
%
%In HPC, FaaS brings low bootup times and flexible resource management to \emph{evolving} and \emph{malleable} %jobs~\cite{10.1007/BFb0022284}, desired traits to scale up in the application phase with adaptive parallelism.

%\section{HPC FaaS Programming Model}
%\input{secs/programming_model}

\section{Case Studies}

\label{sec:evaluation}

We evaluate our HPC software disaggregation approach in three steps: attempting
to answer the following questions:

\begin{enumerate}
  \item How does HPC FaaS perform on a Cray supercomputer?
  \item What is the cost of co-locating functions with batch jobs?
  \item Can disaggregation improve system utilization?
  \item Can HPC applications on a supercomputer benefit from serverless acceleration with rFaaS?
\end{enumerate}
We conduct experiments on two HPC systems:

\paragraph{Ault}
We deploy \toolname{} on cluster nodes with two 18-core Intel Xeon Gold 6154 CPU @ 3.00GHz and 377 GB of memory.
%
%We measure a latency of 3.69 $\mu$s and a bandwidth of 11.69 GiB/s between nodes.
%
We use Docker 20.10.5 with executor image \texttt{ubuntu:20.04},
g++ 10.2, and OpenMPI 4.1.

\paragraph{Daint}
We deploy CPU and GPU co-location jobs on the supercomputing system Piz Daint~\cite{pizDaint}.
The multi-core nodes have two 18-core Intel Xeon E5-2695 v4 @ 2.10GHz
and 128 GB of memory.
The GPU nodes have one 12-core Intel Xeon E5-2690 v3 @ 2.60GHz with 64 GB of memory,
and a NVIDIA Tesla P100  GPU.
All nodes are connected with the Cray Aries interconnect, and we extend
rFaaS with \code{libfabric} to target the uGNI network communication library.
We use Clang 12 and Cray MPICH.

\begin{table}[t!]
  \footnotesize
  \centering
  \begin{adjustbox}{width=\linewidth}
  \begin{tabular}{lcccccccc}
    \toprule
    App /\ Functions  & 1 & 2 & 4 & 8 & 12 & 16 & 24 & 32  \\
    \midrule
    BT, W  & 1 & 1.95 & 3.8 & 6.9 & 9.5 & 11.7 & 17.37 & 23.3  \\
    CG, A  & 1 & 1.85 & 2.8 & 4.8 & 5.8 & 6 & 8.5 & 11.4  \\
    EP, W  & 1 & 2 & 3.78 & 6.8 & 10.2 & 13.6 & 20.4 & 27.2  \\
    LU, W  & 1 & 1.9 & 3.76 & 6.7 & 9.96 & - & 19.7 & -  \\
    \bottomrule
  \end{tabular}
  \end{adjustbox}
  \caption{Relative throughput of an idle-node handling rFaaS functions executing NAS benchmarks.}
  \label{tab:idle_nodes_results}
\end{table}
\begin{figure}[t!]
	\centering
  \includegraphics[width=0.95\linewidth]{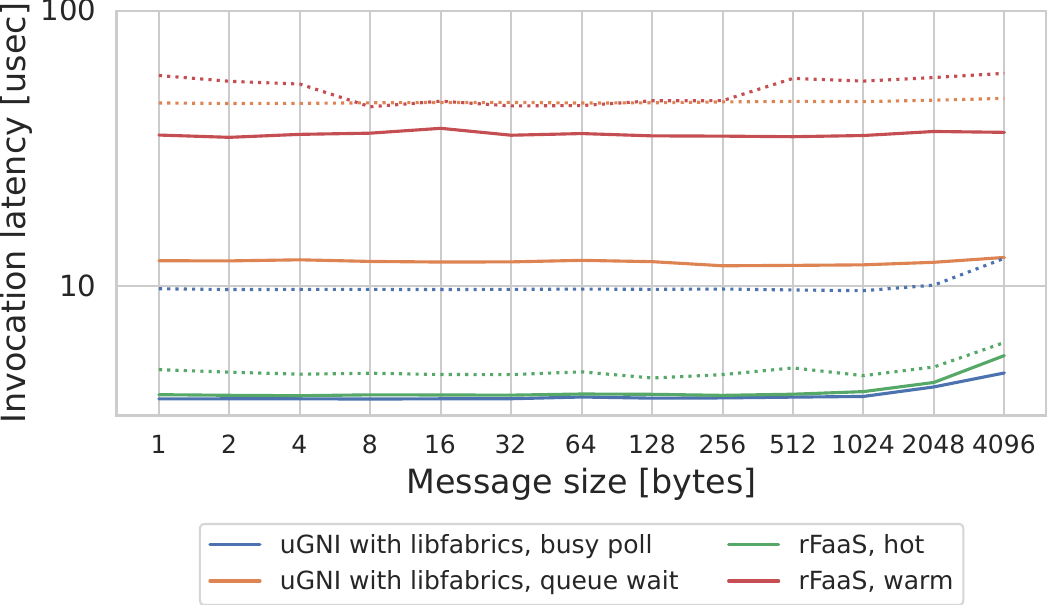}
  \caption{Latency of rFaaS and \emph{libfabric}. The logarithmic plot shows the median (straight) and the 95th percentile (dotted).}
  \label{fig:rfaas_latency}
\end{figure}
\begin{figure}[t!]
	\centering
  \includegraphics[width=0.95\linewidth]{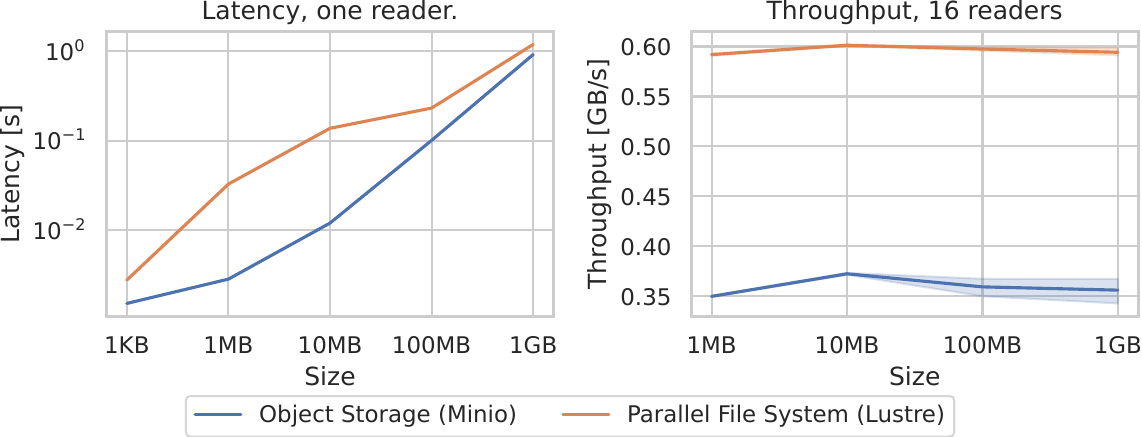}
  \caption{Performance of I/O systems on Piz Daint: Lustre file system versus MinIO object storage.}
  \label{fig:rfaas_io}
\end{figure}

\subsection{rFaaS on Cray Systems}
\label{sec:evaluation_rfaas}
First, we evaluate whether \toolname{} provides the low-latency invocations needed in HPC (Sec.~\ref{sec:faas_interconnect}).
We measure the round-trip time of function invocations on Piz Daint
by using a no-op function with different sizes of input and output data.
We test the \emph{warm} invocations with non-busy waiting methods that have
lower CPU overhead at the cost of increased latency, and the \emph{hot} invocations
that process invocations by continuously polling.
We compare rFaaS against state-of-the-art libfabric benchmarks to check how efficiently we use the network infrastructure 
when serving serverless invocations (Fig.~\ref{fig:rfaas_latency}).
While warm executors need more time to respond,
the hot executions have comparable performance to bare-metal network transport and show consistent performance.

Then, we check if functions can achieve better I/O performance by using the parallel filesystem instead of object storage, a typical solution in FaaS (Sec.~\ref{sec:faas_context}).
To that end, we compare I/O read operations of cloud object storage MinIO~\cite{minio} and the Lustre system on Piz Daint.
By deploying a varying number of readers on different nodes (Fig.~\ref{fig:rfaas_io}),
we show that object storage delivers lower latency for smaller file sizes.
However, Lustre achieves higher throughput at scale.
Thus, replacing cloud storage with a filesystem provides higher I/O performance for HPC functions at no additional cost.

\subsection{Idle Nodes}
We evaluate the effectiveness of using idle nodes for short-running computations.
We select serial NAS benchmarks with runtimes between 0.6 and 4.2 seconds as examples of workloads that can benefit from temporarily idle nodes.
We evaluate the overall throughput of the node when increasing the number of co-located functions (Table~\ref{tab:idle_nodes_results}).
To compute the co-location efficiency, we take the execution with a single rFaaS executor as baseline, and divide the obtained throughput increase by the number of colocated executors.
Except for the CG benchmark, co-located functions achieve 70-80\% efficiency when running simultaneously.
Furthermore, the added overhead of rFaaS execution is around 13\% for the shortest CG, and 
below 1\% on other benchmarks.
Thus, fine-grained and containerized HPC functions can share the node and handle significantly more invocations than an exclusive allocation. 

\begin{figure}[t]
	\centering
  \subfloat[Slowdown of the LULESH batch job.]{
    \includegraphics[width=0.95\linewidth]{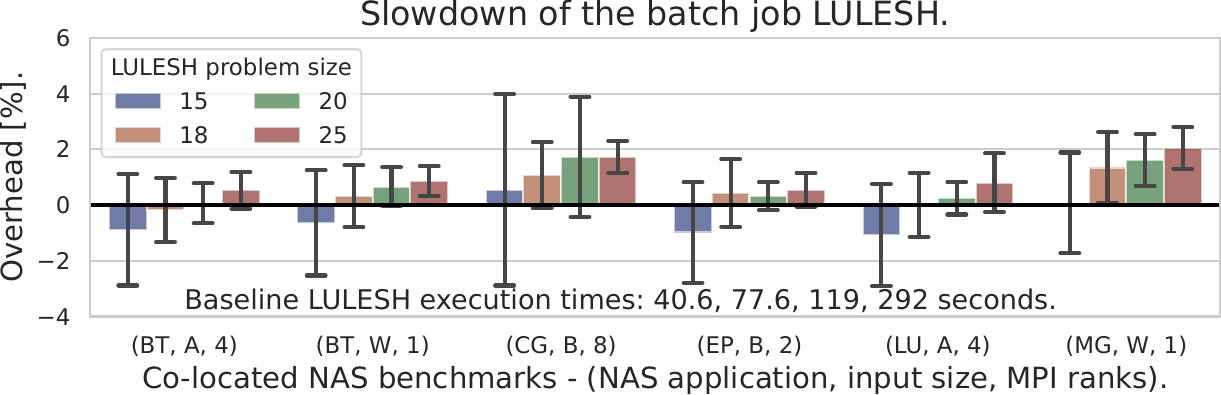}
    \label{fig:colocation_cpu_lulesh}
  }
  \hfill
  \subfloat[Slowdown of the FaaS-like MPI application co-located with LULESH.]{
    \includegraphics[width=0.95\linewidth]{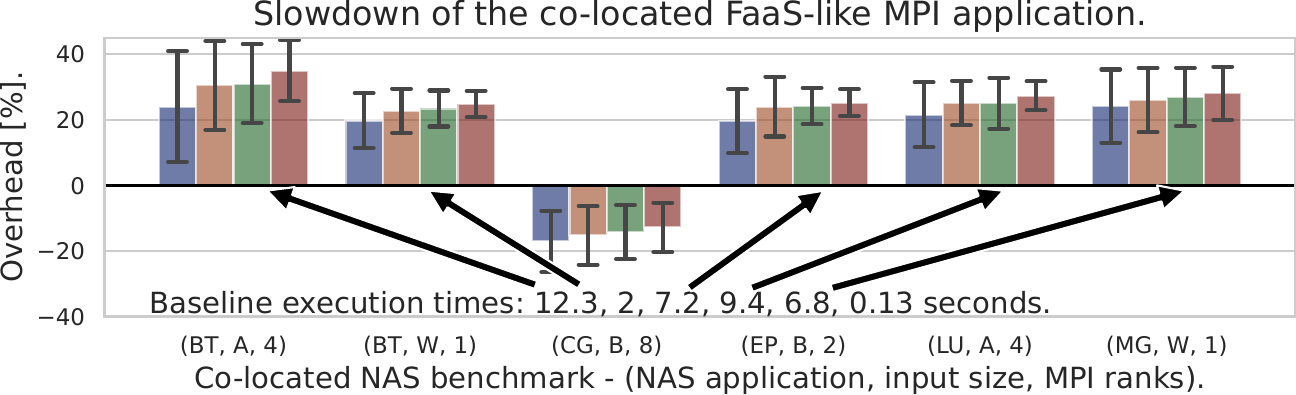}
    \label{fig:colocation_cpu_nas}
  }
  \hfill
  \subfloat[Slowdown of the MILC batch job.]{
    \includegraphics[width=0.95\linewidth]{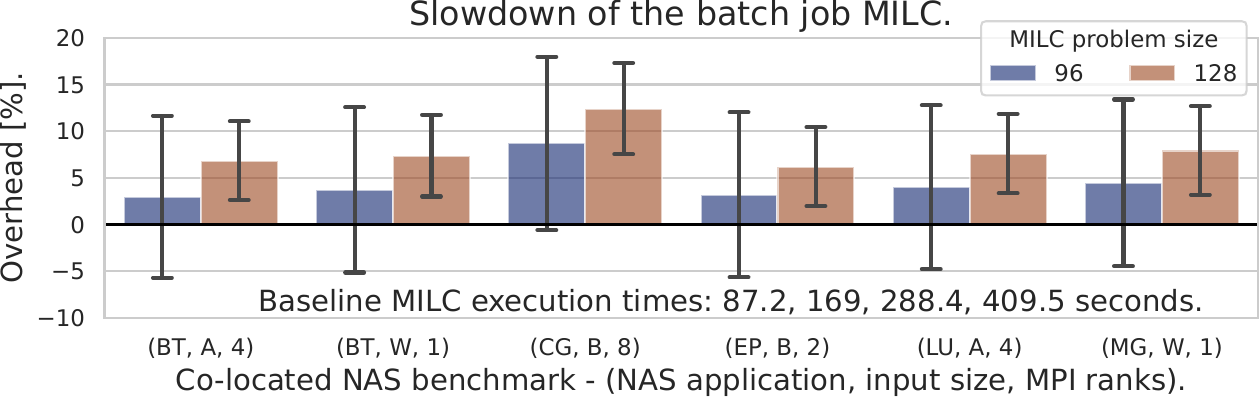}
    \label{fig:colocation_cpu_milc}
  }
  \caption{Overheads of batch jobs co-located with FaaS-like jobs sharing CPUs on idle cores, reported mean with standard deviation over ten repetitions.}
  \label{fig:colocation_cpu}
\end{figure}

\begin{figure}[t]
	\centering
    \includegraphics[width=\linewidth]{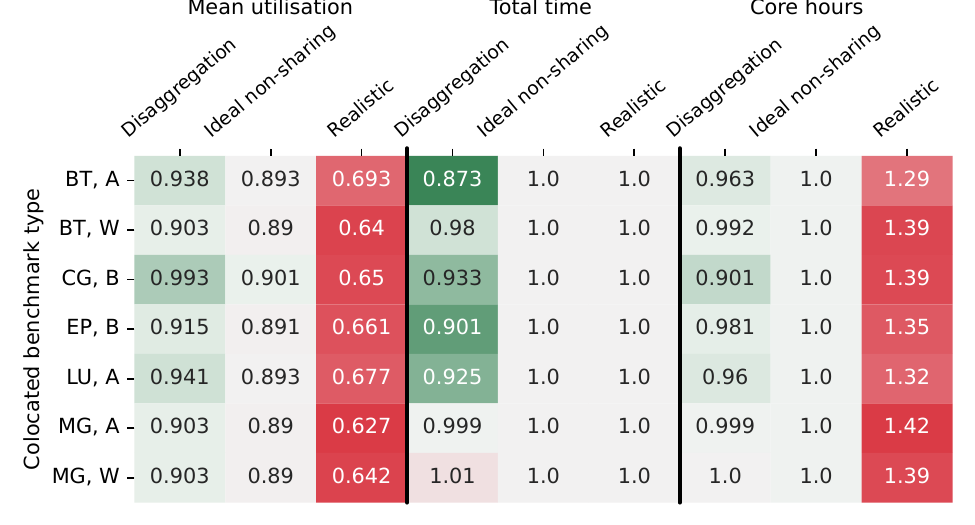}
      \caption{System utilization of co-located execution, a partially co-located execution, and a standard exclusive node allocation.}
      \label{fig:utilization}
\end{figure}

\subsection{Co-location}
\label{sec:evaluation_colocation}

\textbf{ CPU Sharing.}
To evaluate the overhead of co-locating applications by sharing CPUs, we use
the LULESH~\cite{osti_1117905} and MILC~\cite{bernard1991studying}
applications as a classical batch job, using 64 MPI processes and various problem sizes.
We deploy LULESH on 2 Piz Daint nodes, using 32 out of the 36 available cores.
It's important to note that LULESH can only run using a cubic number of processes, e.g., 8, 27, 64, 125, etc. Therefore, using all cores of a node is impossible in many configurations.
Then, we run concurrently NAS benchmarks in the Sarus container on the remaining cores, using CPU binding of tasks through SLURM.
We spread MPI processes equally across two nodes and launch new executions as soon as the previous ones finish.
We chose NAS benchmarks because they are a standard performance indicator~\cite{jin1999openmp}
that represents a variety of compute and communication-bound tasks~\cite{6114174,1592684},
with different memory size~\cite{Shan1900}, data locality and access patterns~\cite{LOFF2021743}, and communication volume~\cite{DBLP:conf/pdcs/FarajY02,1592684}.
Since they have a short runtime that corresponds with execution characteristics of functions~\cite{copik2021sebs,254430,BAUER2024558},
they can represent a FaaS-like workload covering the large diversity of computational patterns that can appear when offloading HPC tasks to serverless.

Fig.~\ref{fig:colocation_cpu} shows that the impact of co-location on the
batch job with this workload is \textbf{negligible},
with changes in LULESH performance explained by the measurement noise. More importantly, only requesting 32 out of 36 cores on each node translates to a core-hour cost reduction of $\approx11\%$, more than offsetting any impact of co-location.
We evaluate the increased system utilization by comparing our co-location with two other scenarios: a realistic exclusive node allocation and an \emph{ideal} allocation where small-scale jobs execute exclusively but are billed for used cores only.
Figure~\ref{fig:utilization} demonstrates significant utilization improvements of up to 52\%.
While the performance loss on the container is higher, it is not a
limitation as HPC functions effectively provide users with a way to
use resources that would otherwise be wasted:
a co-located FaaS-like application is essentially free.

\begin{figure}[t]
  \centering
  \subfloat[LULESH, 27 ranks.]{
    \includegraphics[width=0.95\linewidth]{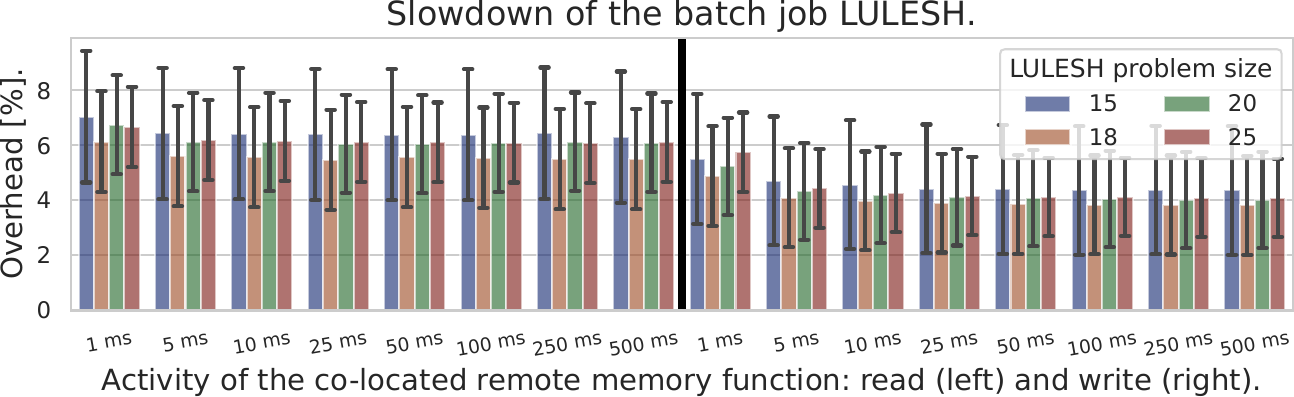}
    \label{fig:colocation_rma_lulesh}
  }
  \hfill
  \subfloat[LULESH, 125 ranks.]{
    \includegraphics[width=0.95\linewidth]{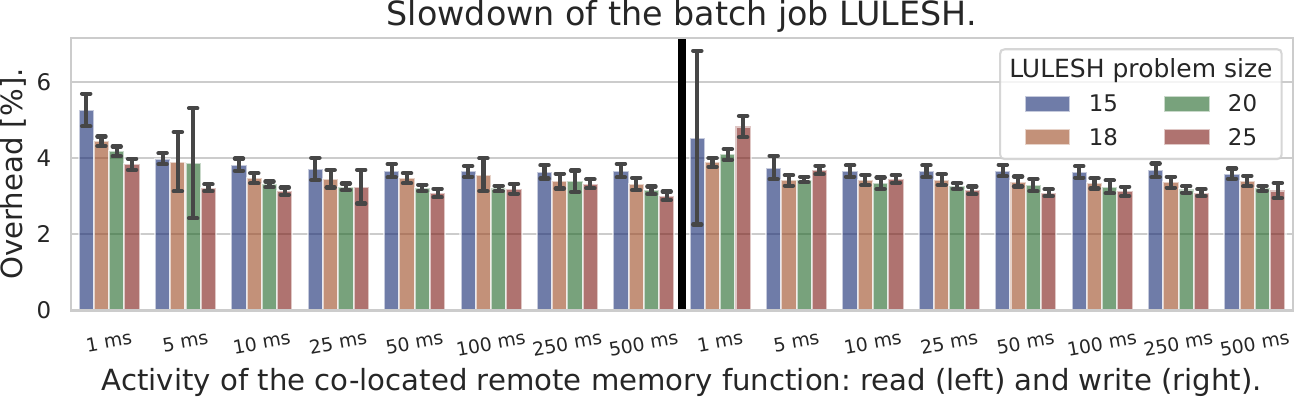}
    \label{fig:colocation_rma_lulesh_125}
  }
  \hfill
  \subfloat[MILC, 32 ranks.]{
    \includegraphics[width=0.95\linewidth]{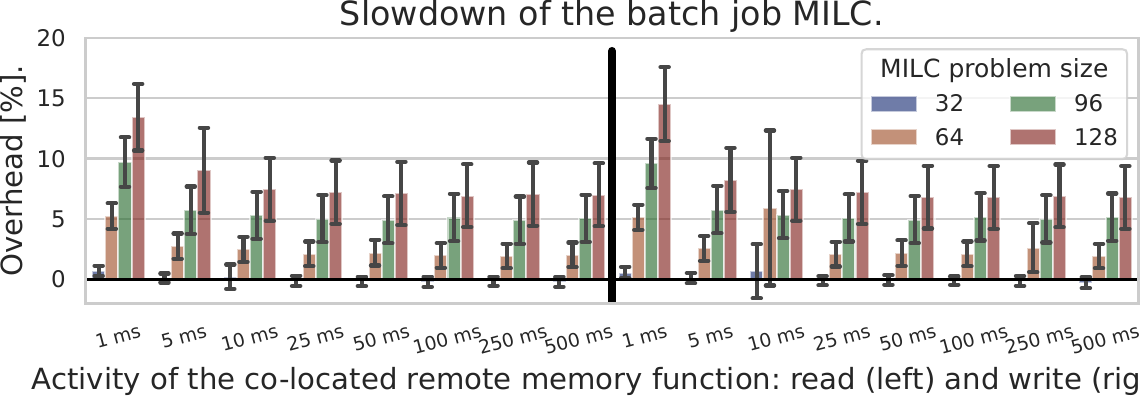}
    \label{fig:colocation_rma_nas}
  }
  \caption{Overhead of batch jobs co-located with \toolname functions providing remote memory. Reported mean with standard deviation over ten repetitions.}
  
  \label{fig:colocation_rma}
\end{figure}

\textbf{Memory Sharing.}
\label{sec:evaluation_remote_memory}
We evaluate the impact of allowing \toolname to use idle memory.
On the Ault system, we run LULESH using 27 and 125 cores, and MILC using 32 cores out of 36 available cores.
We deploy \toolname{}~with the remote memory function setup in a Docker container.
The \toolname{}~function allocates 1 GB of pinned memory available for RDMA operations, and returns the buffer data to the owner.
While running LULESH and MILC, we perform RDMA read and write operations of 10 MB repeatedly with different intervals between repetitions to
test how additional traffic affects performance (Fig.~\ref{fig:colocation_rma}).
The results show that LULESH is not sensitive to the variable perturbation, regardless of problem size, while MILC is more sensitive at larger problem sizes.
When scaling LULESH to multiple nodes, the overall runtime of the job is affected minimally, proving
that compute-intensive applications can share network bandwidth to improve the overall system throughput.
Interestingly, the rate at which data is read or written does not affect performance, even when adding 10GB/s of traffic to the system.
This result is not surprising, as MILC is known the be memory-intensive~\cite{osti_923361,6217478},
and extremely sensitive to both memory bandwidth~\cite{antypas2008nersc,10.1007/978-3-031-40843-4_25}
and to network performance~\cite{antypas2008nersc,smith2016analyzing,9229646,10.1145/3447818.3460362}.
The memory serving function impacts both the available memory bandwidth and the intra-node communication based on shared memory.

\begin{figure}[t]
	\centering
  \subfloat[Slowdown of the LULESH batch job.]{
    \includegraphics[width=0.95\linewidth]{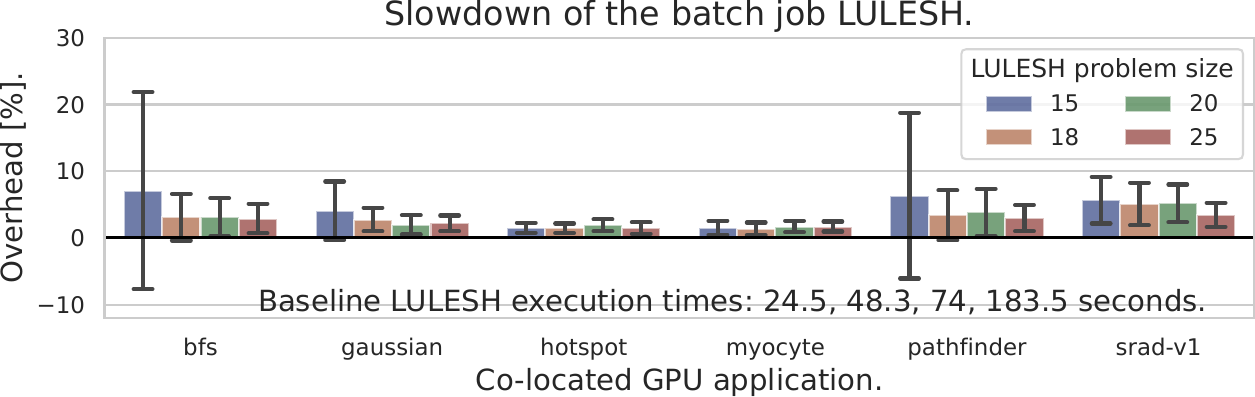}
    \label{fig:colocation_gpu_lulesh}
  }
  \hfill
  \subfloat[Slowdown of the MILC batch job.]{
    \includegraphics[width=0.95\linewidth]{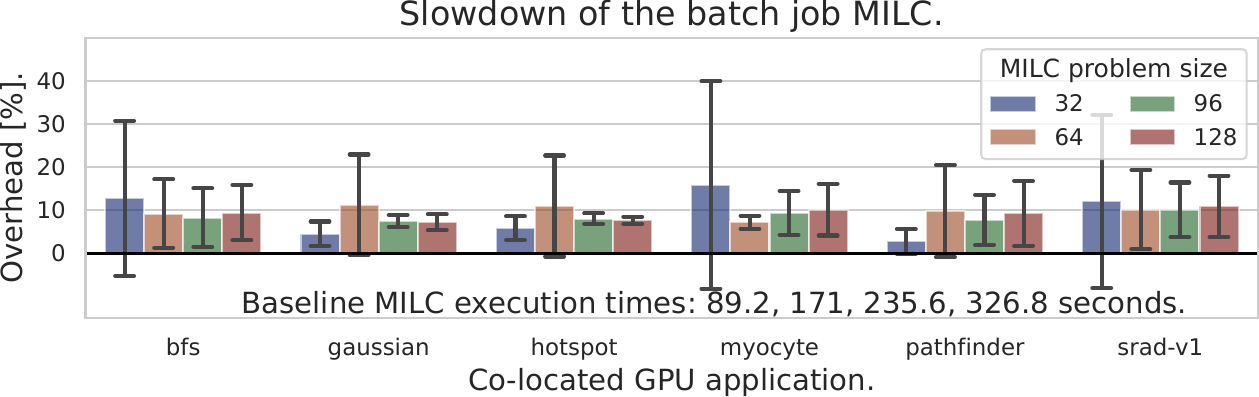}
    \label{fig:colocation_gpu_nas}
  }
  \caption{Overheads of batch jobs sharing node with GPU applications. Reported mean with standard deviation.}
  \label{fig:colocation_gpu}
\end{figure}

\textbf{GPU Sharing.}
\label{sec:evaluation_gpu}
We also run the CPU version\footnote{The original paper states here incorrectly "GPU" instead of "CPU".} of LULESH and MILC on three GPU nodes of the Piz Daint system using 27 ranks and 9 cores out of
the 12 available on each node for LULESH and 32 ranks (divided as 11, 11, and 10 cores) for MILC.
Then,
%On one of the remaining cores,
we run Rodinia GPU benchmarks~\cite{5306797} in a Sarus container (Fig.~\ref{fig:colocation_gpu}), binded to one of the remaining CPU cores through SLURM.
These benchmarks simulate GPU functions as each only takes a few hundred milliseconds.
The overall overhead remains very low ($<5\%$), except for two outliers ($6.1\%$ and $10.5\%$) -- both encountered only for the smallest problem size of LULESH.
However, only requesting 9 out of 12 cores on each GPU node\footnote{The original paper states here incorrectly "GPU" instead of "GPU node".} translates to a core-hour cost reduction of $25\%$, yet again more than offsetting any impact of co-location.
For MILC, the overhead is slightly higher, with the smaller problem sizes experiencing a stronger perturbation.

\begin{bluebox}
Co-location of compute-intensive and memory-bound HPC applications with \toolname{} functions and FaaS-like HPC workloads can be achieved without major overheads in batch jobs, regardless of the shared resource. 
Allocating only required resources reduces batch job costs, even when considering co-location overheads.
\end{bluebox}

\subsection{HPC Integration}
\label{sec:eval_hpc}

\begin{figure}[t]
  \centering
  \subfloat[Black-Scholes method, 100 repetitions.]{
    \includegraphics[width=0.85\linewidth]{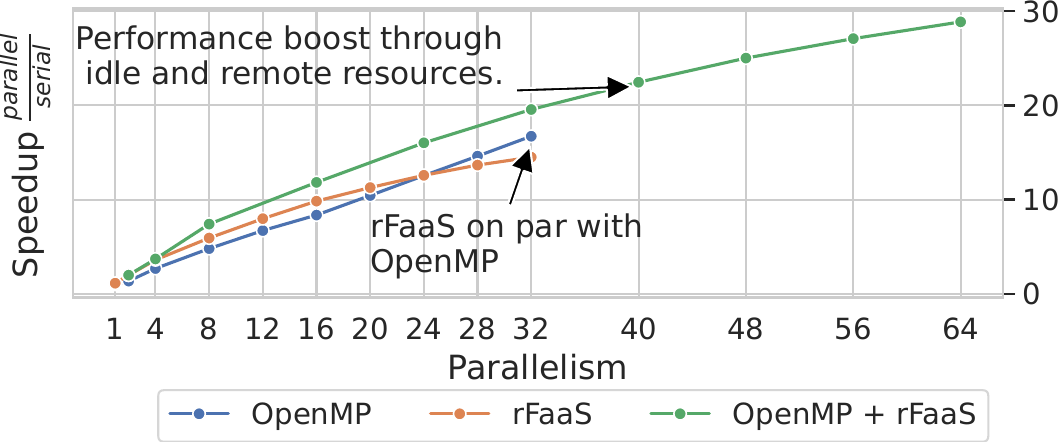}
    \label{fig:res_blackscholes}
  }
  \hfill
  \subfloat[OpenMC, 1,000 particles.]{
    \includegraphics[width=0.85\linewidth]{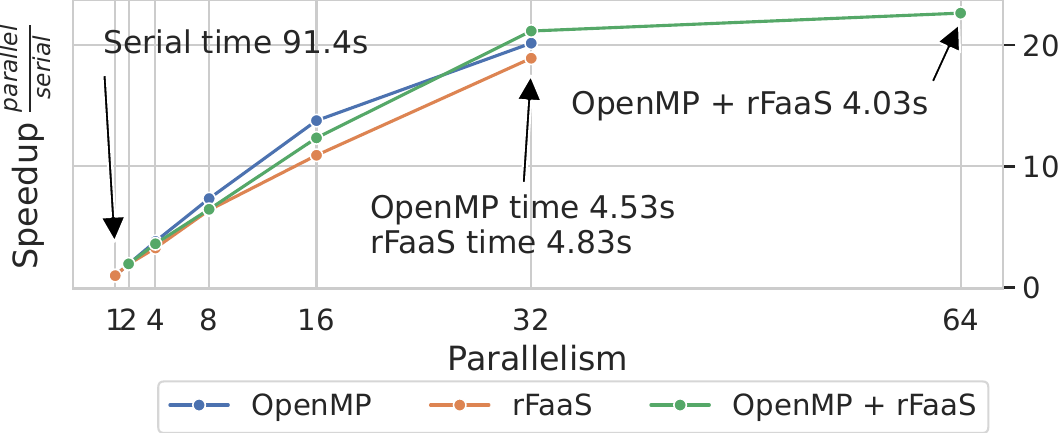}
    \label{fig:res_openmc_1000}
  }
  \hfill
  \subfloat[OpenMC, 10,000 particles.]{
    \includegraphics[width=0.85\linewidth]{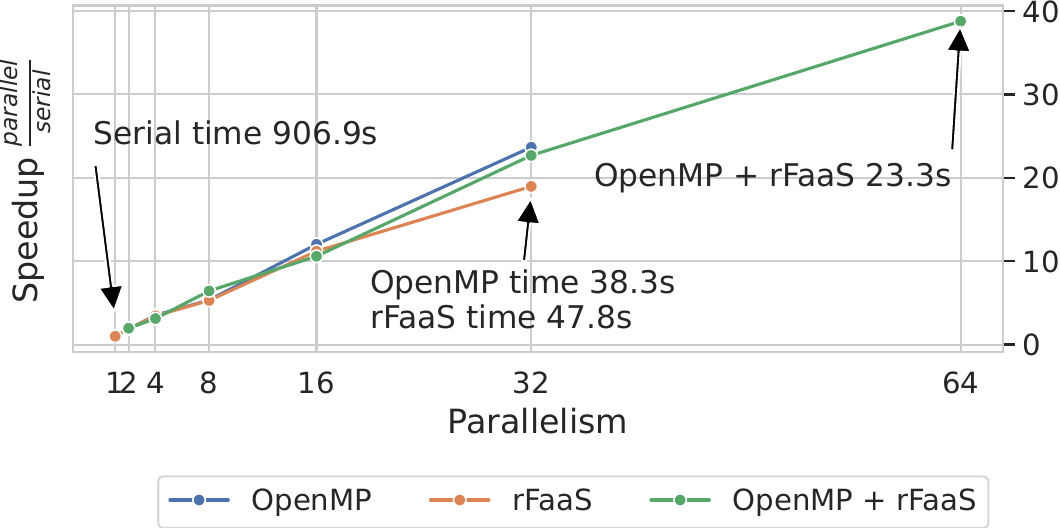}
    \label{fig:res_openmc_10000}
  }
  \caption{\toolname{} in practice, reporting medians.}
\end{figure}

To prove that offloading computations to HPC functions offers performance competitive to native applications, we integrate rFaaS functions into OpenMP benchmarks executed on the Piz Daint and the Ault cluster.
In each application, we move loop code to a separate function, allocate arrays in RDMA-enabled buffers, and replace OpenMP pragmas with \toolname{} dispatch.
We compare the runtimes of benchmarks using OpenMP with runs where the amount of resources has been doubled by allocating one function for each thread.
Thus, we test acceleration by offloading computations to cheap and idle resources
while constrained by the network bandwidth.
This setup allows the dynamic adaptation of parallel allocation and scaling beyond resources available on a single node. 

\subsubsection{Use-case: Black-Scholes simulation}

Figure~\ref{fig:res_blackscholes} demonstrates an OpenMP Black-Scholes benchmark from the PARSEC suite executed with 100 iterations,
modified to use \toolname{} offloading by changing 85 lines of code.
The serial execution takes 726 milliseconds on an input of 229 MB.
We compare the OpenMP version against complete remote execution with \toolname{},
and against doubling parallel resources with cheap serverless allocation.
The application demonstrates efficient offloading of computations until network saturation is reached.

\subsubsection{Use-case: OpenMC}

Figure~\ref{fig:res_openmc_1000} and Figure~\ref{fig:res_openmc_10000} demonstrate OpenMC~\cite{OpenMC}, a Monte Carlo particle transport code %an  %OpenMP Black-Scholes benchmark from the PARSEC suite
modified to use \toolname{} offloading by adding 180 lines of code.
We execute the \emph{opr} benchmark~\cite{OpenMC-oprbenchmark} modeling an Optimized Power Reactor for the input configurations of simulating 1,000 and 10,000 particles on Ault nodes with two AMD EPYC 7742 64-Core Processor @ 2.25GHz and 256 GB memory each. 
In both configurations, functions and clients read 410.8 MB input from the parallel filesystem.
We compare the OpenMP version against complete remote execution with \toolname{}
and against doubling parallel resources with cheap serverless allocation.

\begin{bluebox}
HPC functions can improve massively parallel OpenMP applications, even with millisecond-scale runtime.
\end{bluebox}

\section{Related Work}

Prior work on HPC-oriented serverless targeted scientific applications on a federated platform funcX~\cite{10.1145/3369583.3392683} and HPC workflows~\cite{10046081,10.1145/3503221.3508407}.
We introduce a platform redesigned to match the software and hardware stack of supercomputers and share nodes with batch jobs.
Furthermore, funcX is designed for computation across federated resources, which comes with dozens of milliseconds of invocation latency. 
Our HPC-specialized rFaaS brings microsecond-scale invocations within the same system, efficiently offloading functions running for less than 100-200 ms (Sec.~\ref{sec:eval_hpc}).

\textbf{Resource Underutilization}
Snavely et al.~\cite{https://doi.org/10.1002/cpe.3187} proposed node
sharing with co-location of applications with compatible resource consumption patterns.
However, the detection and avoidance of performance interference
%a major issue and 
requires changes to pricing models~\cite{6877470}, batch systems, and schedulers
~\cite{6877470,frank2019effects,tang2019spread,park2019copart}.
Instead, we propose a decentralized approach with fine-grained functions that does
not require changes in batch systems and online monitoring for interference.
Przybylski et al.~\cite{10046086} proposed to improve HPC system utilization by using idle nodes for serverless platform OpenWhisk.
However, their approach does not consider co-location and fine-grained access to heterogeneous node resources, and uses a generic serverless platform handling cloud workloads.
We define requirements for HPC functions, specialize them to supercomputing environments, and demonstrate integration into HPC applications.
%
%Finally, idle memory can be used to duplicate content for higher throughput~\cite{panwar2019quantifying}.

In the cloud, utilization is improved by harvesting over-allocated virtual machine resources, including CPU cores~\cite{258967} and memory~\cite{10.1145/3503222.3507725}.
Harvested VMs can be used to host invocations of serverless functions~\cite{10.1145/3477132.3483580,10.1145/3503222.3507725}.
Freyr and Libra conduct resource harvesting from over-provisioned serverless functions~\cite{10.1145/3485447.3511979,10.1145/3588195.3592996}.
Our work is focused on HPC resources available for a short time, while the allocation of virtual machines can last for months.
Instead of modifying a classical FaaS platform such as OpenWhisk, we specialize in high-performance serverless systems to HPC systems and applications.

%Finally, idle memory in an HPC system can be used to duplicate memory contents
%for higher throughput~\cite{panwar2019quantifying}.

%rCUDA increases the efficiency of GPU usage in a cluster with remote and virtualized access to GPU devices~\cite{duato2010rcuda},
%minimizing the number of devices in a system.
%
%In contrast, serverless functions move the entire execution context to a remote location.
%
%Backfilling improves system utilization and throughput by reshuffling smaller jobs in a queue to fill utilization gaps~\cite{srinivasan2002characterization}.
%
%Our solution differs in focusing on short-lived functions that operate at a much finer granularity, targeting resources with short availability and supporting multiple classes of resources.
%
%Furthermore, backfilling scheduling could use FaaS to execute jobs in co-location.

\textbf{Elastic MPI}
%
%Adaptive MPI frameworks implement restarting applications
%with different numbers of processes~\cite{6008941},
%reconfiguration frameworks~\cite{MARTIN201560},
%processor virtualization~\cite{10.1145/1122971.1122976},
%and checkpoiting with migration~\cite{10.1007/978-3-319-61982-8_18,10.1007/11752578_32,4215427}.
Adaptive MPI frameworks implement restarting applications~\cite{6008941},
reconfiguration~\cite{MARTIN201560},
processor virtualization~\cite{10.1145/1122971.1122976},
and checkpointing with migration~\cite{10.1007/978-3-319-61982-8_18,10.1007/11752578_32,4215427}.
In contrast, HPC functions bring a dynamic acceleration
with resources allocated on the fly, and require neither restarting nor reconfiguring
the MPI program to incorporate new resources.
Supporting malleable and evolving applications requires changes in schedulers and batch
systems~\cite{6957244,7161531,9406729}, and MPI extensions are needed to extend and shrink the number of processes~\cite{10.1145/2966884.2966917}.
Serverless functions can implement malleable and evolving jobs with high resource availability.

\section{Discussion}

This paper proposes a functionally equivalent alternative to hardware resource disaggregation,
achieved by co-locating a serverless platform with classical HPC batch jobs.
In the following, we discuss several questions our approach raises.

\textbf{How does our solution differ from cloud functions?}
While exploring secure multi-tenancy via serverless techniques is already new
in the context of HPC, we go beyond that: we use co-location only as the starting
point and leverage \toolname{} to allow the different resource subsets to be accessed separately.
Furthermore, unlike the multi-tenant co-location of functions in a cloud, we
focus on providing access to different resource categories in the existing node model
of an HPC data center.

\textbf{What are the limitations imposed by rFaaS?}
The programming model offloads tasks to elastic executors,
similarly to many other serverless approaches to parallel computing~\cite{DBLP:journals/corr/JonasVSR17,273743,9284321}.
Our disaggregation solution relies on network bandwidth to move tasks without significant delays.
Furthermore, HPC applications are adapted to support serverless offloading,
a challenge faced by all applications using FaaS.

\textbf{How does our approach compare to hardware solutions?}
Memory disaggregation needs a software layer for remote paging~\cite{10.1145/3606557.3606563}, which can be fully realized in our solution.
Since we target idle memory that can be reclaimed, we propose swapping and migrations to avoid data loss (Sec.~\ref{sec:colocation_memory}).
However, this limitation does not concern applications using ephemeral memory, e.g., in-memory caching for parallel filesystems.
Our approach uses existing HPC interconnects and avoids additional costs.
There is no penalty for running an unmodified HPC application on
an aggregated system, whereas disaggregation always adds latency to reach remote resources.
Although emerging hardware disaggregation technologies can offer nanosecond-scale
latency for remote memory, 
many high-performance applications benefit from remote memory~\cite{201565,199305,7856562},
indicating that a software-based approach
can offer competitive performance at lower costs.

\textbf{Which applications benefit from co-location?}
We demonstrate on two representative HPC applications that software disaggregation
increases the system's utilization thanks to tolerable performance overheads.
However, co-location has been shown to cause minor slowdowns and increase overall system
throughput in many HPC applications,
including memory-bound and network-sensitive workloads~\cite{https://doi.org/10.48550/arxiv.2204.10768,https://doi.org/10.1002/cpe.3187,frank2019effects,8049000,tang2019spread,xu2023characterizing}.
Job striping and spreading~\cite{tang2019spread,https://doi.org/10.1002/cpe.3187} can be realized in our system due to the reduced costs of under-allocation.

\textbf{How can HPC benefit from short functions?}
Different HPC applications can benefit from dynamically offloading computations to idle nodes.
Major examples include distributed tasking systems such as Dask and Ray~\cite{rocklin2015dask,moritz2018ray,9680456}, malleable and evolving MPI applications~\cite{10.1145/3075564.3075585,8026084},
HPC workflows~\cite{10046081,10.1145/3503221.3508407}, and offloading OpenMP loops to remote devices~\cite{10.1007/978-3-031-15922-0_2,10.1007/978-3-031-07312-0_16}.
By including GPU functions, we can support the rapidly growing space of machine learning inference, a computationally expensive and latency-constrained task.

Serverless requires modifications to serialize inputs and compile function code for FaaS deployment.
This process can be eased with a single-source compiler~\cite{copikcppless}, and our remote memory can support offloading OpenMP applications.
Furthermore, function invocations can be determined by compilers that analyze the parallelism and data movement~\cite{10.1145/3295500.3356173}.

\section{Conclusions}

HPC suffers from underutilization since many systems do not have access to hardware resource disaggregation.
Therefore, we propose a \emph{software disaggregation} approach to efficiently
co-locate long-running batch jobs with serverless functions.
We design targeted FaaS approaches for the three main domains of software disaggregation:
idle processors, memory, and accelerators.
Using a high-performance serverless platform, we demonstrate that targeting idle and partially allocated nodes allows HPC users to benefit from reclaimed resources while
minimizing performance losses, improving system throughput by up to 53\% and supporting remote memory with up to 1GB/s traffic without significant performance overheads from node sharing. 
Finally, we provide users with a path to use the reclaimed resources to accelerate HPC applications.

\section*{Acknowledgment}
This project has received funding from EuroHPC-JU
under grant agreements DEEP-SEA, No 955606, and RED-SEA, No 955776.
This work was partially supported by the ETH Future Computing Laboratory (EFCL), financed by a donation from Huawei Technologies.
Larissa Schmid is supported by the Ministry of Science, Research and the Arts Baden-Württemberg (Az: 7712.14-0821-2) and the pilot program Core Informatics of the Helmholtz Association (HGF).
We would like to thank the Swiss National Supercomputing Centre (CSCS) for providing us with access to their supercomputing infrastructure.

\bibliographystyle{IEEEtran}
\bibliography{serverless,cloud,paper,rdma,hpc}

\end{document}